\documentclass[aps,preprint,amsmath,amssymb,amsfonts]{revtex4}
\usepackage{epsfig}
\usepackage{graphicx}
\usepackage{subfigure}
\usepackage{dcolumn}
\usepackage{bm}
\usepackage{amsthm}
\usepackage{enumitem}
\usepackage{slashed}
\usepackage{braket}
\usepackage{amsmath}
\usepackage{mathtools}

\makeatletter
\def\l@subsubsection#1#2{}
\makeatother

\newcommand{\del}{\partial}

\renewcommand{\Re}{\operatorname{Re}}
\renewcommand{\Im}{\operatorname{Im}}
\newcommand{\sign}{\operatorname{sign}}
\newcommand{\tr}{\operatorname{tr}}

\newcommand{\diag}{\operatorname{diag}}

\newcommand{\bbC}{\mathbb{C}}
\newcommand{\bbI}{\mathbb{I}}

\newcommand{\bbR}{\mathbb{R}}

\newcommand{\calL}{\mathcal{L}}

\newcommand{\scri}{\mathcal{I}}

\begin{document}

\title{Scattering in the static patch of de Sitter space}

\author{Emil Albrychiewicz}
\email{ealbrych@berkeley.edu}
\affiliation{University of California, Berkeley, CA 94720, U.S.A.}
\author{Yasha Neiman}
\email{yashula@icloud.com}
\affiliation{Okinawa Institute of Science and Technology, 1919-1 Tancha, Onna-son, Okinawa 904-0495, Japan}

\date{\today}

\begin{abstract}
We study the scattering problem in the static patch of de Sitter space, i.e. the problem of field evolution between the past and future horizons of a de Sitter observer. We calculate the leading-order scattering for a conformally massless scalar with cubic interaction, as both the simplest case and a warmup towards Yang-Mills and gravity. Our strategy is to decompose the static-patch evolution problem into a pair of more symmetric evolution problems in two Poincare patches, sewn together by a spatial inversion. To carry this out explicitly, we end up developing formulas for the momentum-space effect of inversions in flat spacetime. The geometric construction of an electron's 4-momentum and spin vectors from a Dirac spinor turns out to be surprisingly relevant. 
\end{abstract}

\maketitle
\tableofcontents
\newpage

\section{Introduction} \label{sec:intro}

\subsection{Why scattering in the de Sitter static patch?} \label{sec:intro:why}

In field theory and gravity, a special role is played by scattering problems, or, more generally, by observables defined at the asymptotic boundary of spacetime. Of course, one reason for this is mathematical physics for its own sake: boundary observables are worth exploring, simply becase they form a well-defined and relatively simple subset of all possible questions in field theory. Another reason is that sometimes an asymptotic quantity turns out to be directly relevant to some experimental or observational setup. The obvious case is that of scattering amplitudes in flat spacetime, which directly describe collider experiments. Similar hopes are now being placed on future-boundary correlators in (nearly) de Sitter spacetime: within the paradigm of inflation, these should become observable as non-Gaussianities in the Cosmic Microwave Background \cite{Maldacena:2002vr}. Fundamentally, though, the main reason to be interested in asymptotic correlations is that currently these are the only observables we can make sense of in quantum gravity. 

There is just one uncomfortable detail: as far as we can tell from observations, our Universe is undergoing an accelerated expansion driven by a positive cosmological constant, which should lead to de Sitter asymptotics in the future. In such a spacetime, observers such as ourselves are trapped inside their cosmological horizons, without causal access to asymptotic infinity. This leaves two possibilities. One is that the naive extrapolation into the distant future is wrong, and that the Universe's present de Sitter-like phase is merely temporary, just like the earlier de Sitter phase that is conjectured in inflation (see e.g. \cite{Obied:2018sgi}; for a review of the difficulties with de Sitter space in string theory, see \cite{Danielsson:2018ztv}). The other possibility is that we take our de Sitter fate seriously. We must then grapple with the question of how to think about quantum gravity within a finite region of space, without observables at infinity. This appears truly daunting. In fact, a plausible reading of the theoretical evidence is that it can't be done without modifying Quantum Mechanics itself. For what it's worth, our lot is thrown with the position that this direction is ultimately the correct one, and that it signals the next revolution in fundamental theory.

\begin{figure}%
	\centering%
	\includegraphics[scale=.8]{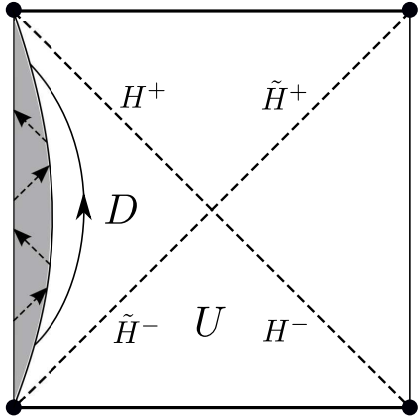} \\
	\caption{A Penrose diagram of $dS_4$, with a cartoon observer inside the static patch $D$. The shaded area is the observer's ``body'' -- a collection of worldlines that maintain causal contact through the exchange of signals at or below the speed of light. Exiting and re-entering the shaded area is the worldline of a ``probe'' dispatched by the observer to gather some data from outside his body. The static patch $D$ is the largest spacetime region that can be covered by such an observer. Its boundaries are the past horizon $\tilde H^-$ and future horizon $H^+$. The region $U$, while inside the observer's past lightcone, isn't really observable. For instance, the observer may measure the fields on $\tilde H^-$, but the fields within $U$ can't be deduced from these without some assumptions about unobservable data e.g. on $H^-$.}
	\label{fig:observer} 
\end{figure}

In the meantime, the least we can do is to familiarize ourselves with the available observables in an asymptotically de Sitter universe. For simplicity, we leave real-world cosmology aside, and focus on a pure de Sitter spacetime $dS_4$. We also leave aside the complications of dynamical geometry, and focus on either non-gravitational field theory, or gravity viewed as perturbations over a $dS_4$ background. Further, we wish to retain as much contact as possible with the familiar realm of asymptotic observables. What, then, is the ``most asymptotic'' observable in $dS_4$? Clearly, this will be an observable defined not at the boundary of spacetime, but at the boundary of the largest observable region. This largest region is a \emph{static patch} of $dS_4$ -- the patch contained between the past and future cosmological horizons of a bulk observer, depicted in figure \ref{fig:observer}.

This last statement is actually the subject of frequent confusion, which we should address before proceeding. It is often stated that an observer can see the entire spacetime region inside his past lightcone. This certainly \emph{seems} true from both everyday and scientific experience: aren't we seeing the Andromeda galaxy when we point our binoculars in its direction? However, our ability to think this way is conditioned on an orderly, i.e. low-entropy, state of the Universe. From the point of view of \emph{fundamental} physics, we've never seen the Andromeda galaxy: we are just measuring the electromagnetic field on our retina, and perhaps on various nearby surfaces, e.g. inside telescopes. An honest observation/measurement is always local, and honest inferences from them are always restricted to the causal domain of dependence of the measured region. It is in this strict, field-theoretic sense that the largest observable region of $dS_4$ is the static patch. This is illustrated from a different point of view in figure \ref{fig:observer}, and discussed further in \cite{Halpern:2015zia}.

With that off our chest, let us concentrate on the static patch. Its boundary is a pair of lightlike cosmological horizons, one in the past and one in the future, much like the null past/future boundary of Minkowski space. The natural observable would thus be akin to the Minkowski S-matrix: an evolution from initial data on the past horizon to final data on the future horizon. What do we mean by ``data''? In classical field theory, these would just be the values of the fields on each horizon (because the horizons are lightlike, the fields' normal derivative doesn't need to be specified separately). This statement of the problem can be carried over to the quantum level, where we would now seek to express the \emph{field operators} on the future horizon in terms of those on the past horizon. Note that this isn't quite the path that's usually taken in the flat-spacetime case: there, one tends to talk about scattering \emph{amplitudes} between initial and final \emph{states}. Of course, if the evolution of field operators is known, then the evolution of states can be derived from it by acting with the operators on a vacuum state. This vacuum state will usually be annihilated by half of the field operators, specifically those with negative frequency. In this paper, we will prefer to deal with fields rather than states. This is for two reasons. First, the most well-behaved vacuum state in $dS_4$ is the Bunch-Davies vacuum, but in the static patch this corresponds to a thermal state, not a pure one. Second, we'll find it useful to work with a combination of positive-frequency and negative-frequency modes in the Bunch-Davies sense, rather than restrict to one or the other. Summing up, then, our general problem statement will be to express the fields on the final horizon in terms of those on the initial horizon.

\subsection{Scope and structure of the paper}

To our knowledge, there is virtually no published work on static-patch scattering. This is probably for two reasons. The first is that this problem has no claim to direct observational relevance (unlike e.g. the future boundary correlators of the inflationary $dS_4$). The second is its low degree of symmetry. The spacetime symmetry of the $dS_4$ static patch is a meager $\bbR\times SO(3)$. The $SO(3)$ decribes spatial rotations, while the $\bbR$ describes time translations in the static patch; in Poincare coordinates, the latter take the form of dilatations. Particularly painful in its absence is a spatial translation symmetry, which would allow us to work in momentum space. A central message of the present paper is that there are ways around this low degree of symmetry. In particular, we can construct the static-patch evolution by first evolving the fields from the initial horizon to the $dS_4$ boundary, and then from the boundary back to the final horizon. Each of these evolutions is taking place in a \emph{Poincare} patch of $dS_4$. The latter has spatial translation symmetry, which allows us to work in momentum space, performing essentially the same calculations as in the standard framework of cosmological correlators. Each of the two Poincare patches is conformal to (half of) Minkowski spacetime, and the two patches are related to each other by a spacetime inversion (which reduces to a spatial inversion on the boundary). As a result, our main technical task becomes expressing the effects of spatial and spacetime inversions in momentum space.

In the present paper, we will apply this strategy to a particular simple field theory in $dS_4$, at leading order in the interaction. Specifically, we consider a conformally massless scalar field with cubic coupling. For this theory, we calculate the evolution from initial to final horizon at tree-level (i.e. classically), to quadratic order (i.e. with just a single cubic vertex). This work is a sequel to \cite{David:2019mos}, where the non-interacting version of the problem was considered for massless fields of all spins. In this paper, we will ignore ``soft'' effects. Specifically, we will assume an input field configuration that vanishes at the initial horizon's edges (i.e. at the intersections with the past boundary and with the future horizon), and we will calculate the output fields up to terms concentrated on the edges of the final horizon.

The rest of the paper is structured as follows. In section \ref{sec:root}, we set up our momentum-space framework, along with the decomposition of the static-patch problem into Poincare-patch problems. In section \ref{sec:M}, we perform a standard calculation of the Poincare-patch evolution. In section \ref{sec:inversion}, we derive the necessary formulas for spacetime inversions, using spinor techniques. In section \ref{sec:result}, we present the final result for the static-patch scattering. The result includes a partial cancellation between the two Poincare-patch evolutions. As we will see, this is a general feature that will occur whenever a Poincare-patch amplitude happens to have the full $SO(1,4)$ de Sitter symmetry. In the simple case of single-vertex diagrams, this will in turn occur whenever the Poincare patch amplitude does not have singularities on the energy axis. In section \ref{sec:discuss}, we present this argument in more detail, along with its expected consequences for Yang-Mills and GR, and other closing remarks.

For the reader in a hurry, the key formulas are as follows. Our encoding of the field data on the initial and final horizons is given in eqs. \eqref{eq:horizon_value_final_even},\eqref{eq:horizon_value_initial_even}. The scattering is given by eqs. \eqref{eq:S_even_general}-\eqref{eq:S_even}. These make use of the Minkowski-space inversion kernels from section \ref{sec:inversion}, which are summarized in eqs. \eqref{eq:I_1_summary_result}-\eqref{eq:A}.

\section{Decomposition into Poincare-patch evolutions} \label{sec:root}

The field theory that we'll consider in this paper is that of a conformally massless scalar with cubic coupling. The theory lives in $dS_4$ spacetime, whose curvature radius we set to 1. The Lagrangian is:
\begin{align}
 \calL = -\frac{1}{2}g^{\mu\nu}\del_\mu\varphi\del_\nu\varphi - \varphi^2 - \frac{\alpha}{3}\varphi^3 \ , \label{eq:L}
\end{align}
where $\alpha$ is the coupling constant, and $g_{\mu\nu}$ is the $dS_4$ metric. We won't be bothered by the fact that the $\varphi^3$ potential is unbounded from below -- this theory is just a toy example, and it is healthy enough perturbatively (or classically). The field equation for the Lagrangian \eqref{eq:L} reads:
\begin{align}
(\Box - 2)\varphi = \alpha\varphi^2 \ , \label{eq:field_eq}
\end{align}
where $\Box$ is the $dS_4$ d'Alembertian. In this section, we will show how the static-patch scattering problem decomposes into a pair of Poincare-patch problems, using the theory \eqref{eq:L} as an example.

\subsection{Geometric framework} \label{sec:root:geometry}

We define de Sitter space $dS_4$ as the sphere of unit spacelike radius in 5d Minkowski space $\bbR^{1,4}$. We use lightcone coordinates $(u,v,\mathbf{r})$ for $\bbR^{1,4}$, such that its metric is $ds^2 = -dudv + \mathbf{dr}^2$. The $dS_4$ spacetime is then defined by the 4d hypersurface $-uv + \mathbf{r}^2 = 1$. The conformal boundary of $dS_4$ is described by the limit:
\begin{align}
 (u,v,\mathbf{r}) \ \rightarrow \ \frac{1}{t}(U, V, \mathbf{R}) \ , \label{eq:bdry_limit}
\end{align}
with $t\rightarrow 0$ and $(U, V, \mathbf{R})$ a null vector in $\bbR^{1,4}$, i.e. $-UV + \mathbf{R}^2 = 0$. This description is redundant under simulataneous rescalings of $t$ and $(U, V, \mathbf{R})$ by the same finite factor; these correspond to local rescalings of the boundary metric. The conformal boundary is composed of two 3-spheres: the 3-sphere $\scri^+$ of future-pointing null directions in $\bbR^{1,4}$, and the sphere $\scri^-$ of past-pointing ones. The notation $t$ for the prefactor in \eqref{eq:bdry_limit} is not a common one, but it will make the discussion below less cumbersome; in bulk Poincare coordinates, $t$ will become the conformal time. 

The 3d hypersurfaces $(u=0,\mathbf{r}^2 = 1)$ and $(v=0,\mathbf{r}^2 = 1)$ form a pair of cosmological horizons in $dS_4$. We will refer to them respectively as the initial horizon $\tilde H$ and the final horizon $H$. Each of these horizons is a 2-sphere's worth of lightrays: the unit vector $\mathbf{r}$ specifies our position on the 2-sphere, and the lightlike coordinate $u$ or $v$ specifies our position along the lightrays. The $v<0$ past half $\tilde H^-$ of $\tilde H$ and the $u>0$ future half $H^+$ of $H$ form the past and future boundaries of a static patch. The symmetries of the static patch are the $SO(3)$ rotations of $\mathbf{r}$ and the ``time translations'' $(u,v)\rightarrow (e^\tau u, e^{-\tau} v)$, which are actually boosts in the $uv$ plane of $\bbR^{1,4}$.

The static-patch scattering problem is to find the final field configuration $\varphi(u,\mathbf{r})$ on $H^+$ (with $u>0$ and $\mathbf{r}^2 = 1$) as a functional of the initial field configuration $\varphi(v,\mathbf{r})$ on $\tilde H^-$ (with $v<0$ and $\mathbf{r}^2 = 1$). Our strategy will be to work instead with the full horizons $\tilde H,H$. That is, we will extend the initial data on $\tilde H^-$ to all of $\tilde H$, evolve that to $H$, and then restrict to $H^+$. In addition, we will break down the evolution from $\tilde H$ to $H$ into two steps: we will first evolve from $\tilde H$ backwards in time to the conformal boundary $\scri^-$, and then evolve forward to $H$. The advantage here is that each of these evolutions takes place in a \emph{Poincare patch} of $dS_4$, with its relatively high degree of symmetry. Instead of $\scri^-$ as the intermediate hypersurface, we could alternatively use $\scri^+$, which would be more in line with the cosmological literature. The choice is ultimately arbitrary; we will stick here with $\scri^-$, since it will lead to fewer minus signs along the way.

Let us now make the above discussion more explicit. We define Poincare coordinates $x^\mu = (t,\mathbf{x})$ for $dS_4$ associated with the final horizon $H$, and Poincare coordinates $\tilde x^\mu = (\tilde t,\mathbf{\tilde x})$ associated with the initial horizon $\tilde H$. These are related to the embedding-space coordinates $(u,v,\mathbf{r})$ as:
\begin{align}
 (u,v,\mathbf{r}) = \frac{1}{t}\left(t^2 - \mathbf{x}^2, -1, \mathbf{x}\right) = \frac{1}{\tilde t}\left(-1, \tilde t^2 - \mathbf{\tilde x}^2, \mathbf{\tilde x}\right) \ . \label{eq:Poincare}
\end{align}
The $x^\mu = (t,\mathbf{x})$, with $t>0$, span the Poincare patch to the past of the horizon $H$. The same coordinates with $t<0$ span the patch to the future of $H$, which we will mostly ignore; note that $t\rightarrow -t$ implements the antipodal map $(u,v,\mathbf{r})\rightarrow -(u,v,\mathbf{r})$ in $dS_4$. The limit $t\rightarrow 0^+$ describes the past conformal boundary $\scri^-$, which is then coordinatized by $\mathbf{x}$, via the limiting procedure \eqref{eq:bdry_limit} with $(U,V,\mathbf{R}) = (-\mathbf{x}^2,-1,\mathbf{x})$. The horizon $H$ is described by the simultaneous limit $t,|\mathbf{x}|\rightarrow \infty$, with $t-|\mathbf{x}|$ kept finite. The horizon coordinates $(u,\mathbf{r})$ are then related to $(t,\mathbf{x})$ via:
\begin{align}
 u = 2(t-|\mathbf{x}|) \ ; \quad \mathbf{r} = \frac{\mathbf{x}}{|\mathbf{x}|} \ . \label{eq:horizon_final}
\end{align}
The evolution from $\scri^-$ to $H$ can now be viewed as evolution in the Poincare patch $t>0$, all the way from $t=0^+$ to $t=\infty$. The same remarks apply to the Poincare coordinates $\tilde x^\mu = (\tilde z,\tilde{\mathbf{x}})$, for which $\tilde t=0^+$ is again the past boundary $\scri^-$, while the horizon $\tilde H$ is given by the limit $\tilde t,|\mathbf{\tilde x}|\rightarrow \infty$ with:
\begin{align}
  v = 2(\tilde t-|\mathbf{\tilde x}|) \ ; \quad \mathbf{r} = \frac{\mathbf{\tilde x}}{|\mathbf{\tilde x}|} \ . \label{eq:horizon_initial}
\end{align}

The $dS_4$ metric in the Poincare coordinates takes the conformally flat form:
\begin{align}
 \begin{split}
   ds^2 &= \frac{-dt^2 + \mathbf{dx}^2}{t^2} = \frac{1}{t^2}\,\eta_{\mu\nu}dx^\mu dx^\nu \\
     &= \frac{-d\tilde t^2 + \mathbf{d\tilde x}^2}{\tilde t^2} = \frac{1}{\tilde t^2}\,\eta_{\mu\nu}d\tilde x^\mu d\tilde x^\nu \ , 
 \end{split} \label{eq:Poincare_metric}
\end{align}
where $\eta_{\mu\nu}$ is the 4d Minkowski metric. From the point of view of this metric, the boundary $t=0$ is just an ordinary time slice, while the horizon \eqref{eq:horizon_final} or \eqref{eq:horizon_initial} is future lightlike infinity. 

We can read off from \eqref{eq:Poincare} the relationship between the $x^\mu$ and $\tilde x^\mu$ coordinates as:
\begin{align}
 \tilde x^\mu = (\tilde t,\mathbf{\tilde x}) = \frac{(t,\mathbf{x})}{\mathbf{x}^2 - t^2} = \frac{x^\mu}{\eta_{\nu\rho}x^\nu x^\rho} \ . \label{eq:inversion}
\end{align}
This is simply an inversion in Minkowski spacetime. In particular, on the boundary $t=\tilde t = 0$, the two frames are related by a spatial inversion $\mathbf{\tilde x} = \mathbf{x}/\mathbf{x}^2$.

Our decomposition of the static-patch problem thus takes the form:
\begin{align}
 S &= R\hat S E \ ; \label{eq:reduction_1} \\ 
 \hat S &= MIM^{-1} \ . \label{eq:reduction_2}
\end{align}
Here, $S$ is the desired evolution matrix from $\tilde H^-$ to $H^+$. Eq. \eqref{eq:reduction_1} describes the relatively trivial step of replacing it by an evolution matrix $\hat S$ from all of $\tilde H$ to all of $H$. Specifically, $E$ is an extension (which we are free to choose) of the initial data on $\tilde H^-$ onto all of $\tilde H$, while $R$ is the reduction of the final data on $H$ onto $H^+$. The less trivial step is eq. \eqref{eq:reduction_2}. There, $M$ stands for the evolution in a Poincare patch, from the conformal boundary $t=0$ to the horizon $t=\infty$, while $I$ represents a coordinate inversion \eqref{eq:inversion}. Eq. \eqref{eq:reduction_2} then decomposes the $\tilde H\rightarrow H$ evolution into $\tilde H\rightarrow\scri^-$ (described by $M^{-1}$), followed by a switch from one Poincare frame to another (described by $I$), and finally an evolution from $\scri^-$ to $H$ (described by $M$). Note that the only step here that depends on the dynamics (i.e. on the interaction) is the Poincare-patch evolution $M$: the rest is kinematical. In this sense, eqs. \eqref{eq:reduction_1}-\eqref{eq:reduction_2} constitute a complete \emph{solution} of static-patch scattering in terms of Poincare-patch evolution. That being said, the inversion operation $I$, though kinematical, is not quite trivial to perform, and will make up the subject of section \ref{sec:inversion}.

We close this subsection with some comments on the structure of eq. \eqref{eq:reduction_2}. First, let us note its similarity to the standard procedure \cite{Maldacena:2002vr,Maldacena:2011nz} for calculating cosmological correlators at $\scri^+$: there, one also evolves from the horizon to the boundary, and then back to the horizon. The differences are:
\begin{enumerate}
 \item In \cite{Maldacena:2002vr,Maldacena:2011nz}, one imposes the Bunch-Davies vacuum on the horizon, whereas we are evolving general initial fields into final fields.
 \item In \cite{Maldacena:2002vr,Maldacena:2011nz}, one inserts operators at the boundary, whereas we instead perform an inversion there.
\end{enumerate}
Finally, let us comment on the spacetime symmetries of the Poincare-patch evolution $M$. By construction, it has the 3d translation, rotation and dilatation symmetries of the Poincare patch. However, this isn't the end of the story: as we'll discuss in section \ref{sec:discuss}, certain pieces of $M$ have the full $SO(1,4)$ symmetry of de Sitter space, i.e. full 3d conformal symmetry. Such pieces of $M$ will commute with the inversion $I$ in \eqref{eq:reduction_2}, and cancel with their counterparts in $M^{-1}$. Due to such cancellations, the static-patch scattering $S$ ends up in some sense simpler than the Poincare-patch evolution $M$, despite having lower overall symmetry. Problems that have a square root, as in \eqref{eq:reduction_2}, always carry some pleasant surprises!

\subsection{Plane waves in the Poincare patch} \label{sec:root:waves}

Let us now solve the linearized field equation $(\Box - 2)\varphi = 0$, in the Poincare coordinates associated with e.g. the final horizon $H$. The differential operator $\Box - 2$ is conformal to the flat d'Alembertian $\eta^{\mu\nu}\del_\mu\del_\nu$, with the field rescaled by the conformal factor of $t$:
\begin{align}
 (\Box - 2)\varphi = t^3\eta^{\mu\nu}\del_\mu\del_\nu(t^{-1}\varphi) \ . 
\end{align}
Therefore, the general solution to the free equation is simply $t$ times a superposition of plane waves, parameterized by spatial momentum $\mathbf{p}$ and by the sign of the energy:
\begin{align}
 \varphi(t,\mathbf{x}) = t\int\frac{d^3\mathbf{p}}{2|\mathbf{p}|}\left(a(\mathbf{p})e^{i(\mathbf{p\cdot x} - |\mathbf{p}|t)} + a^\dagger(\mathbf{p})e^{-i(\mathbf{p\cdot x} - |\mathbf{p}|t)} \right) \ . \label{eq:wave_basis}
\end{align}
Our chosen measure $\frac{d^3\mathbf{p}}{2|\mathbf{p}|}$ over momentum space is the Lorentz-invariant measure over lightlike momenta in the 4d Minkowski space defined by $x^\mu = (t,\mathbf{x})$. In the quantum picture, the coefficients $a(\mathbf{p})$ and $a^\dagger(\mathbf{p})$ describe annihilation and creation operators over the Bunch-Davies vacuum. We can write the solution \eqref{eq:wave_basis} more compactly by unifying $a(\mathbf{p})$ and $a^\dagger(\mathbf{p})$ into a single function $a(p_\mu)$ of a lightlike 4-momentum $p_\mu = (p_t,\mathbf{p})$, which can be either future-pointing (positive energy, $p_t=-|\mathbf{p}|$) or past-pointing (negative energy, $p_t=+|\mathbf{p}|$). In particular, we denote $a(\mathbf{p})\equiv a(-|\mathbf{p}|,\mathbf{p})$ and $a^\dagger(-\mathbf{p})\equiv a(|\mathbf{p}|,\mathbf{p})$. The solution \eqref{eq:wave_basis} then takes the form:
\begin{align}
 \varphi(x^\mu) = t\int_{p^2 = 0} \frac{d^3\mathbf{p}}{2|\mathbf{p}|}\, a(p_\mu)\,e^{ip_\mu x^\mu} \ , \label{eq:wave_basis_covariant} 
\end{align}
where $\int_{p^2 = 0}$ is a shorthand for integration over both halves of the lightcone $p_t = \pm |\mathbf{p}|$:
\begin{align}
 \int_{p^2 = 0} \equiv \int_{p_t = |\mathbf{p}|} + \int_{p_t = -|\mathbf{p}|} \ .
\end{align}
We will also find it useful to decompose the free field \eqref{eq:wave_basis} into even and odd pieces under the antipodal map $t\rightarrow -t$:
\begin{align}
 \varphi_{\text{odd}}(t,\mathbf{x}) &= t\int\frac{d^3\mathbf{p}}{2|\mathbf{p}|}\,b(\mathbf{p})e^{i\mathbf{p\cdot x}}\cos(|\mathbf{p}|t) \ ; \quad b(\mathbf{p}) = a(\mathbf{p}) + a^\dagger(-\mathbf{p}) \ ; \\
 \varphi_{\text{even}}(t,\mathbf{x}) &= t\int\frac{d^3\mathbf{p}}{2|\mathbf{p}|}\,c(\mathbf{p})e^{i\mathbf{p\cdot x}}\sin(|\mathbf{p}|t) \ ;  \quad c(\mathbf{p}) = -i\left(a(\mathbf{p}) - a^\dagger(-\mathbf{p})\right) \ .
\end{align}
These have different $t$ scalings near the conformal boundary $t=0$:
\begin{align}
 \varphi_{\text{odd}}(t,\mathbf{x}) \, &\longrightarrow \, t\int\frac{d^3\mathbf{p}}{2|\mathbf{p}|}\,b(\mathbf{p})e^{i\mathbf{p\cdot x}} \ ; \\ 
 \varphi_{\text{even}}(t,\mathbf{x}) \, &\longrightarrow \, \frac{t^2}{2}\int d^3\mathbf{p}\,c(\mathbf{p})e^{i\mathbf{p\cdot x}} \ .
\end{align}
In other words, $\varphi_{\text{odd}}$ and $\varphi_{\text{even}}$ have boundary conformal weights $\Delta = 1$ and $\Delta = 2$, respectively.

Also important is the behavior of the plane waves \eqref{eq:wave_basis} at the horizon. Up to the conformal rescaling by $t$, this is directly analogous to the asymptotics of plane waves at Minkowski lightlike infinity. In particular, a wavepacket with small but nonzero spread around a mean momentum $\mathbf{p}$ will end up focusing in the direction of $\mathbf{p}$. To see this explicitly, let's decompose the momentum integral in \eqref{eq:wave_basis} into an integral over the magnitude $\omega\equiv |\mathbf{p}|$ and the direction $\mathbf{\hat p} \equiv \mathbf{p}/|\mathbf{p}|$:
\begin{align}
  \varphi(t,\mathbf{x}) = \frac{t}{2}\int_0^\infty \omega d\omega \left(e^{-i\omega t}\int_{S_2} d^2\mathbf{\hat p}\,a(\omega\mathbf{\hat p})\,e^{i\omega|\mathbf{x}|\mathbf{\hat p\cdot\hat x}} + c.c. \right) \ ,
\end{align}
where we denoted $\mathbf{\hat x}\equiv \mathbf{x}/|\mathbf{x}|$. In the horizon limit, $|\mathbf{x}|$ becomes very large. The exponential $e^{i\omega|\mathbf{x}|\mathbf{\hat p\cdot\hat x}}$ is then a rapidly oscillating phase, and the $d^2\mathbf{\hat p}$ integral can be performed in the stationary-phase approximation. The stationary points are $\mathbf{\hat p} = \pm \mathbf{\hat x}$, where the exponential takes the values $e^{\pm i\omega|\mathbf{x}|}$. These combine with the $e^{-i\omega t}$ factor to give $e^{i\omega(-t\pm |\mathbf{x}|)}$. For the lower sign choice, this is again a rapidly oscillating phase, which will be killed by the $d\omega$ integral. Thus, only the $\mathbf{\hat p} = \mathbf{\hat x}$ stationary point contributes. The Hessian of the phase $\omega|\mathbf{x}|\mathbf{\hat p\cdot\hat x}$ at this point (in an orthonormal basis on $S_2$) is $-\frac{1}{2}\omega|\mathbf{x}|\diag(1,1)$. The $d^2\mathbf{\hat p}$ integral can now be evaluated as:
\begin{align}
 \varphi(t,\mathbf{x}) \,\longrightarrow\, \frac{\pi t}{|\mathbf{x}|}\int_0^\infty d\omega\left(-ia(\omega\mathbf{\hat x}) e^{i\omega(|\mathbf{x}| - t)} + c.c. \right) \ .
\end{align}
In the horizon limit, we have $t/|\mathbf{x}|\rightarrow 1$, while $|\mathbf{x}| - t$ and $\mathbf{\hat x}$ become horizon coordinates according to \eqref{eq:horizon_final}. We thus obtain the value of the free solution \eqref{eq:wave_basis} at a horizon point $(u,\mathbf{r})$ as:
\begin{align}
 \varphi(u,\mathbf{r}) &= -\pi i\int_0^\infty d\omega\left(a(\omega\mathbf{r}) e^{-i\omega u/2} - a^\dagger(\omega\mathbf{r}) e^{i\omega u/2} \right) \ ,
\end{align}
which decomposes into antipodally even and odd parts as:
\begin{align}
 \varphi_{\text{odd}}(u,\mathbf{r}) &= -\frac{\pi i}{2}\int_0^\infty d\omega\left(b(\omega\mathbf{r}) e^{-i\omega u/2} - b(-\omega\mathbf{r}) e^{i\omega u/2} \right) \ ; \label{eq:horizon_value_final_odd} \\
 \varphi_{\text{even}}(u,\mathbf{r}) &= \frac{\pi}{2}\int_0^\infty d\omega\left(c(\omega\mathbf{r}) e^{-i\omega u/2} + c(-\omega\mathbf{r}) e^{i\omega u/2} \right) \ .  \label{eq:horizon_value_final_even}
\end{align}
Of course, the same plane-wave decomposition can be performed in the Poincare patch $(\tilde t,\mathbf{\tilde x})$ associated with the initial horizon $\tilde H$. We again define the field and its antipodally even/odd parts as:
\begin{align}
   \varphi(\tilde t,\mathbf{\tilde x}) &= \tilde t\int\frac{d^3\mathbf{\tilde p}}{2|\mathbf{\tilde p}|}\left(\tilde a(\mathbf{\tilde p})e^{i(\mathbf{\tilde p\cdot\tilde x} - |\mathbf{\tilde p}|\tilde t)} + c.c. \right)
    = \tilde t \int_{\tilde p^2 = 0} \frac{d^3\mathbf{\tilde p}}{2|\mathbf{\tilde p}|}\,a(\tilde p_\mu)\,e^{i\tilde p_\mu \tilde x^\mu} \ ; \\
 \varphi_{\text{odd}}(\tilde t,\mathbf{\tilde x}) &= \tilde t\int\frac{d^3\mathbf{\tilde p}}{2|\mathbf{\tilde p}|}\,\tilde b(\mathbf{\tilde p})e^{i\mathbf{\tilde p\cdot\tilde x}}\cos(|\mathbf{\tilde p}|\tilde t) \ ; \quad 
  \tilde b(\mathbf{p}) = \tilde a(\mathbf{\tilde p}) + \tilde a^\dagger(-\mathbf{p}) \ ; \\
 \varphi_{\text{even}}(\tilde t,\mathbf{\tilde x}) &= \tilde t\int\frac{d^3\mathbf{\tilde p}}{2|\mathbf{\tilde p}|}\,\tilde c(\mathbf{\tilde p})e^{i\mathbf{\tilde p\cdot\tilde x}}\sin(|\mathbf{\tilde p}|\tilde t) \ ;  \quad 
   \tilde c(\mathbf{p}) = -i\left(\tilde a(\mathbf{\tilde p}) - \tilde a^\dagger(-\mathbf{\tilde p})\right) \ ,
\end{align}
with boundary asymptotics at $\tilde t\rightarrow 0$:
\begin{align}
\varphi_{\text{odd}}(\tilde t,\mathbf{\tilde x}) \, &\longrightarrow \, \tilde t\int\frac{d^3\mathbf{\tilde p}}{2|\mathbf{\tilde p}|}\,\tilde b(\mathbf{\tilde p})e^{i\mathbf{\tilde p\cdot\tilde x}} \ ; \label{eq:boundary_value_initial_odd} \\ 
\varphi_{\text{even}}(\tilde t,\mathbf{\tilde x}) \, &\longrightarrow \, \frac{\tilde t^2}{2}\int d^3\mathbf{\tilde p}\,\tilde c(\mathbf{\tilde p})e^{i\mathbf{\tilde p\cdot\tilde x}} \ , \label{eq:boundary_value_initial_even}
\end{align}
and horizon values:
\begin{align}
 \varphi(v,\mathbf{r}) &= -\pi i\int_0^\infty d\tilde\omega\left(\tilde a(\tilde\omega\mathbf{r}) e^{-i\tilde\omega v/2} - \tilde a^\dagger(\tilde\omega\mathbf{r}) e^{i\tilde\omega v/2} \right) \ ; \label{eq:horizon_value_initial} \\
 \varphi_{\text{odd}}(v,\mathbf{r}) &= -\frac{\pi i}{2}\int_0^\infty d\tilde\omega\left(\tilde b(\tilde\omega\mathbf{r}) e^{-i\tilde\omega v/2} - \tilde b(-\tilde\omega\mathbf{r}) e^{i\tilde\omega v/2} \right) \ ; \\
 \varphi_{\text{even}}(v,\mathbf{r}) &= \frac{\pi}{2}\int_0^\infty d\tilde\omega\left(\tilde c(\tilde\omega\mathbf{r}) e^{-i\tilde\omega v/2} + \tilde c(-\tilde\omega\mathbf{r}) e^{i\tilde\omega v/2} \right) \ .  \label{eq:horizon_value_initial_even}
\end{align}

\subsection{Static-patch scattering in the plane-wave basis}

Our decomposition \eqref{eq:reduction_1}-\eqref{eq:reduction_2} of the static-patch scattering can now be made more explicit. We begin by choosing an extension of the initial data on $\tilde H^-$, i.e. at $v<0$, to all of $\tilde H$. As we will see, the choice that will avoid inconsistencies in our method is the antipodally even one $\varphi(v,\mathbf{r}) = \varphi(-v,-\mathbf{r}) = \varphi_\text{even}(v,\mathbf{r})$. We then Fourier-transform with respect to $v$ as in \eqref{eq:horizon_value_initial_even}, and encode the initial data in terms of plane-wave coefficients $\tilde c_{\tilde H}(\mathbf{\tilde p})$. The non-linear Poincare-patch evolution $M^{-1}$ then evolves the field onto the boundary $\scri^-$, where it has the asymptotic structure \eqref{eq:boundary_value_initial_odd}-\eqref{eq:boundary_value_initial_even} of a free field, but with some new plane-wave coefficients $\tilde b_\scri(\mathbf{\tilde p}),\tilde c_\scri(\mathbf{\tilde p})$. We then apply an inversion $I$ to transform these into plane-wave coefficients $b_\scri(\mathbf{p}),c_\scri(\mathbf{p})$ in the Poincare patch of the final horizon $H$. Finally, we evolve those with the Poincare-patch evolution $M$ into plane-wave coefficients $b_H(\mathbf{p}),c_H(\mathbf{p})$ on $H$ itself, from which we can read off the field on $H$ (and on $H^+$ in particular) via the Fourier transforms \eqref{eq:horizon_value_final_odd}-\eqref{eq:horizon_value_final_even}.

Hidden in the above procedure is one crucial assumption: that near the boundary $\scri^-$, the interacting field can be approximated as free, and therefore has asymptotic behavior of the form \eqref{eq:boundary_value_initial_odd}-\eqref{eq:boundary_value_initial_even}. For this to be true, the non-linear $\varphi^2$ term in the field equation \eqref{eq:field_eq} should vanish at $t\rightarrow 0$ faster than either of the free-field solutions \eqref{eq:boundary_value_initial_odd}-\eqref{eq:boundary_value_initial_even}, i.e. faster than $t^2$. This fails if the $\varphi^2$ term contains two odd factors $\varphi_{\text{odd}}\sim t$, but holds if at least one of the factors is $\varphi_{\text{even}}\sim t^2$. It is for this reason that we choose even initial data on $\tilde H$.

That being said, during the intermediate stages of the calculation, it will be more convenient to work in terms of unrestricted plane-wave coefficients $a(\mathbf{p}),a^\dagger(\mathbf{p})$. The problem with non-even data will then reveal itself as a divergence at small $t$, which we will regularize; upon restricting to even data, the divergence will cancel. For compactness, we again unify $a(\mathbf{p})$ and $a^\dagger(\mathbf{p})$ into a single function $a(p_\mu)$, as in \eqref{eq:wave_basis_covariant}. In this language, the most general evolution $\hat S$ from $\tilde H$ to $H$ (to first order in the interaction, i.e. to second order in the fields) takes the form:
\begin{align}
 \begin{split}
   a_H(k_\mu) ={}& \int_{\substack{\tilde k^2\,=\,0 \\ k\cdot\tilde k\,<\,0}} \frac{d^3\mathbf{\tilde k}}{2|\mathbf{\tilde k}|}\,\bbI_1(k_\mu,\tilde k_\mu)\,\tilde a_{\tilde H}(\tilde k_\mu) \\
    &+ \int_{\tilde p^2 = 0} \frac{d^3\mathbf{\tilde p}}{2|\mathbf{\tilde p}|} \int_{\tilde q^2 = 0} \frac{d^3\mathbf{\tilde q}}{2|\mathbf{\tilde q}|}\,S_2(k_\mu;\tilde p_\mu,\tilde q_\mu)\,\tilde a_{\tilde H}(\tilde p_\mu)\,\tilde a_{\tilde H}(\tilde q_\mu) \ ,
 \end{split} \label{eq:S_general}
\end{align}
i.e. the scattering is parameterized by a function $\mathbb{I}_1(k_\mu,\tilde k_\mu)$ at linear order, and by another function $S_2(k_\mu;\tilde p_\mu,\tilde q_\mu)$ at quadratic order (the font choice in $\bbI_1$ is to distinguish it from Bessel functions below). The final 4-momentum $k_\mu$ in \eqref{eq:S_general} is lightlike, and the integral in the first line is over lightlike 4-momenta $\tilde k_\mu$ with the same time-orientation as $k_\mu$:
\begin{align}
 \int_{\substack{\tilde k^2\,=\,0 \\ k\cdot\tilde k\,<\,0}} \equiv \int_{\tilde k_t = \sign(k_t)|\mathbf{\tilde k}|} \ .
\end{align}
Now, to understand the structure of each term in \eqref{eq:S_general}, let us recall the decomposition \eqref{eq:reduction_2} of the horizon$\rightarrow$horizon evolution into Poincare-patch evolutions. In the plane-wave basis, the Poincare-patch evolution $M$ at leading order is just the identity. Therefore, the linear scattering function $\bbI_1(k_\mu,\tilde k_\mu)$ in \eqref{eq:S_general} is simply the inversion $I$ from \eqref{eq:reduction_2}, which transforms a plane wave $t e^{ip_\mu x^\mu}$ into a superposition of plane waves in the inverted Poincare frame \eqref{eq:inversion}. As we can see from \eqref{eq:inversion}, under inversion, the $t$ factor in the plane wave $te^{ip_\mu x^\mu}$ simply rescales by the conformal factor $\eta_{\mu\nu}x^\mu x^\nu$. Therefore, we can identify $\bbI_1(k_\mu,\tilde k_\mu)$ as the inversion kernel for lightlike plane waves in \emph{flat} spacetime, with conformal weight 1: 
\begin{align}
 \frac{e^{i\tilde p_\mu x^\mu/(\eta_{\nu\rho}x^\nu x^\rho)}}{\eta_{\nu\rho}x^\nu x^\rho} = \int_{\substack{p^2\,=\,0 \\ p\cdot\tilde p\,<\,0}} \frac{d^3\mathbf{p}}{2|\mathbf{p}|}\,\bbI_1(p_\mu,\tilde p_\mu)\,e^{ip_\mu x^\mu} \ . \label{eq:I_1} 
\end{align}
Equivalently, we can evaluate \eqref{eq:I_1} at $t=0$, and think of $\bbI_1(p_\mu,\tilde p_\mu) \equiv \bbI_1(\mathbf{p},\mathbf{\tilde p})$ as the inversion kernel for momenta in $\bbR^3$. We will calculate $\bbI_1$ explicitly in section \ref{sec:inversion}. 

Let's now consider the quadratic term $S_2(k_\mu;\tilde p_\mu,\tilde q_\mu)$ in \eqref{eq:S_general}. To evaluate it, we will need the boundary$\rightarrow$ horizon evolution $M$ and its inverse $M^{-1}$. The most general form of $M$ to quadratic order, evolving the field from $\scri^-$ to $H$, reads:
\begin{align}
 a_H(k_\mu) = a_\scri(k_\mu) + \int_{p^2=0} \frac{d^3\mathbf{p}}{2|\mathbf{p}|} \int_{q^2=0} \frac{d^3\mathbf{q}}{2|\mathbf{q}|}\,M_2(k_\mu;p_\mu,q_\mu)\,a_\scri(p_\mu)\,a_\scri(q_\mu) \ , \label{eq:M_general}
\end{align}
where $M_2(k_\mu;p_\mu,q_\mu)$ is a function of three lightlike 4-momenta -- two incoming $p_\mu,q_\mu$, and one outgoing $k_\mu$. We will calculate this function in section \ref{sec:M}. For now, we note that by spatial translation symmetry, it must contain a momentum-conserving delta function:
\begin{align}
 M_2(k_\mu;p_\mu,q_\mu) = \hat M_2(k_\mu;p_\mu,q_\mu)\,\delta^3(\mathbf{p + q - k}) \ . \label{eq:M_delta}
\end{align}
The inverse evolution $M^{-1}$ from $\tilde H$ to $\scri^-$ takes the same form as \eqref{eq:M_general}, but with the sign of the quadratic term flipped:
\begin{align}
 \tilde a_\scri(\tilde k_\mu) = \tilde a_{\tilde H}(\tilde k_\mu) 
   - \int_{\tilde p^2=0} \frac{d^3\mathbf{\tilde p}}{2|\mathbf{\tilde p}|} \int_{\tilde q^2=0} \frac{d^3\mathbf{\tilde q}}{2|\mathbf{\tilde q}|}\,M_2(\tilde k_\mu;\tilde p_\mu,\tilde q_\mu)\,\tilde a_{\tilde H}(\tilde p_\mu)\,\tilde a_{\tilde H}(\tilde q_\mu) \ .
\end{align}
Sewing the two evolutions together with the inversion \eqref{eq:inversion}, we obtain the quadratic scattering function $S_2(k_\mu;\tilde p_\mu,\tilde q_\mu)$ as:
\begin{align}
 \begin{split}
   S_2(k_\mu;\tilde p_\mu,\tilde q_\mu) ={}& - \int_{\substack{\tilde k^2\,=\,0 \\ k\cdot\tilde k\,<\,0}} \frac{d^3\mathbf{\tilde k}}{2|\mathbf{\tilde k}|}\,\bbI_1(k_\mu,\tilde k_\mu)\,M_2(\tilde k_\mu;\tilde p_\mu,\tilde q_\mu) \\
    &+ \int_{\substack{p^2\,=\,0 \\ p\cdot\tilde p\,<\,0}} \frac{d^3\mathbf{p}}{2|\mathbf{p}|}\,\bbI_1(p_\mu,\tilde p_\mu) \int_{\substack{q^2\,=\,0 \\ q\cdot\tilde q\,<\,0}} \frac{d^3\mathbf{q}}{2|\mathbf{q}|}\,\bbI_1(q_\mu,\tilde q_\mu)\,M_2(k_\mu;p_\mu,q_\mu) \ .
 \end{split} \label{eq:S2_raw}
\end{align} 
The first term in \eqref{eq:S2_raw} is relatively simple, because the momentum-conserving delta function $\delta^3(\mathbf{\tilde p + \tilde q - \tilde k})$ inside $M_2(\tilde k_\mu;\tilde p_\mu,\tilde q_\mu)$ will cancel the $d^3\mathbf{\tilde k}$ integral. The second term is more problematic, because the integral there is over 6 momentum components $d^3\mathbf{p}\,d^3\mathbf{q}$. To simplify it, we make the following observation. Due to the non-derivative form of the $\varphi^3$ coupling, $M_2(k_\mu;p_\mu,q_\mu)$ only depends on the local product of the incoming waves:
\begin{align}
(te^{ip_\mu x^\mu})(te^{iq_\mu x^\mu}) = t^2 e^{iP_\mu x^\mu} \ ; \quad P_\mu \equiv p_\mu + q_\mu \ , \label{eq:P}
\end{align}
where the total 4-momentum $P_\mu \equiv p_\mu + q_\mu$ is now generic, rather than lightlike. We can thus write our evolution functions more compactly as:
\begin{align}
 M_2(k_\mu;p_\mu,q_\mu) &\equiv M_2(k_\mu;P_\mu) =  \hat M_2(k_\mu;P_\mu)\,\delta^3(\mathbf{P - k}) \ ; \label{eq:M_P} \\
 S_2(k_\mu;\tilde p_\mu,\tilde q_\mu) &\equiv S_2(k_\mu;\tilde P_\mu) \ .
\end{align}
More importantly, the double inversion in \eqref{eq:S2_raw} can now be replaced by a single one:
\begin{align}
  S_2(k_\mu;\tilde P_\mu) = -\int_{\substack{\tilde k^2\,=\,0 \\ k\cdot\tilde k\,<\,0}} \frac{d^3\mathbf{\tilde k}}{2|\mathbf{\tilde k}|}\,\bbI_1(k_\mu,\tilde k_\mu)\,M(\tilde k_\mu;\tilde P_\mu) + \int d^4P\,\bbI_2(P_\mu,\tilde P_\mu)\,M(k_\mu;P_\mu) \ , \label{eq:S2}
\end{align}
where $\bbI_2(P_\mu,\tilde P_\mu)$ is the inversion kernel for plane waves \eqref{eq:P} with generic 4-momentum and conformal weight 2:
\begin{align}
  \frac{e^{i\tilde P_\mu x^\mu/(\eta_{\nu\rho}x^\nu x^\rho)}}{(\eta_{\nu\rho}x^\nu x^\rho)^2} = \int d^4P\,\bbI_2(P_\mu,\tilde P_\mu)\,e^{iP_\mu x^\mu} \ . \label{eq:I_2} 
\end{align}
The second term in \eqref{eq:S2} now contains an integral over just 4 momentum components $P_\mu$. 3 of these integrals will be cancelled by the momentum-conserving delta function inside $M(k_\mu;P_\mu)$, leaving just an integral over the energy $P_t$. This simplifies the scattering function \eqref{eq:S2} into:
\begin{align}
 \begin{split}
   S_2(k_\mu;\tilde P_\mu) ={}& -\left.\bbI_1(k_\mu,\tilde k_\mu)\,\hat M_2(\tilde k_\mu;\tilde P_\mu)\right|_{\tilde k_\mu\,=\,(\sign(k_t)|\mathbf{\tilde P}|,\,\mathbf{\tilde P})} \\
     &+ \int dP_t\left.\bbI_2(P_\mu,\tilde P_\mu)\,\hat M_2(k_\mu;P_\mu)\right|_{\mathbf{P} = \mathbf{k}} \ .
 \end{split} \label{eq:S2_integrated}
\end{align}
What remains now is to calculate the Poincare-patch evolution function $\hat M_2(k_\mu,P_\mu)$, and the inversion kernels $\bbI_1(k_\mu,\tilde k_\mu)$ and $\bbI_2(P_\mu,\tilde P_\mu)$. This will be the job of sections \ref{sec:M} and \ref{sec:inversion}, respectively.

\section{Evolution in the Poincare patch} \label{sec:M}

Let us now calculate the Poincare-patch evolution \eqref{eq:M_general} from the boundary $\scri^-$ to the horizon $H$, to first order in the interaction. Since we are at tree-level, this can be done by simply evolving from $t=0$ to $t=\infty$ with the field equation \eqref{eq:field_eq}. We encode the initial data at $t=0$ in terms of plane-wave coefficients $a_\scri(p_\mu)$. The linearized approximation to the bulk solution is given by the free field \eqref{eq:wave_basis}:
\begin{align}
 \varphi^{(1)}(t,\mathbf{x}) = \int d^3\mathbf{p}\,f(t,\mathbf{p})e^{i\mathbf{p\cdot x}} \ ; \quad f(t,\mathbf{p}) = \frac{t}{2|\mathbf{p}|}\left(a_\scri(\mathbf{p})e^{-i|\mathbf{p}|t} + a^\dagger_\scri(-\mathbf{p})e^{i|\mathbf{p}|t} \right) \ , \label{eq:linearized}
\end{align}
where $f(t,\mathbf{p})$ simply denotes the spatial Fourier transform of $\varphi^{(1)}(t,\mathbf{x})$. The quadratic correction to the field \eqref{eq:linearized} due to the non-linear term in the field equation \eqref{eq:field_eq} can be constructed as:
\begin{align}
 \varphi^{(2)}(t,\mathbf{x}) = \alpha\int d^3\mathbf{k}\,e^{i\mathbf{k\cdot x}}\int d^3\mathbf{p} \int_0^\infty dt' f(t',\mathbf{p})f(t',\mathbf{k-p})\,G(t,\mathbf{x};t',\mathbf{k}) \ . \label{eq:quadratic}
\end{align}
Here, $t'$ denotes the time at which the interaction takes place, while $G(t,\mathbf{x};t',\mathbf{k})$ is the Green's function for evolution from $t'$ to a later time $t$, defined as a retarded solution to the linearized field equation with source:
\begin{align}
(\Box - 2)G = t^3\eta^{\mu\nu}\del_\mu\del_\nu(t^{-1} G) = \delta(t-t')e^{i\mathbf{k}\cdot\mathbf{x}} \ ; \quad G = 0 \quad \forall t<t' \ . \label{eq:G_conditions}
\end{align}
The solution to \eqref{eq:G_conditions} is a step function $\theta(t-t')$ multiplied by a combination of plane waves, which is determined by requiring a vanishing value and a normalized $\del_t$ derivative at $t=t'$:
\begin{align}
 G(t,\mathbf{x};t',\mathbf{k}) = -\frac{t}{|\mathbf{k}|t'^3}\,e^{i\mathbf{k\cdot x}} \sin\!\big[|\mathbf{k}|(t-t')\big]\,\theta(t-t') \ . \label{eq:G}
\end{align}
The evolution function $M_2(k_\mu;P_\mu)$ from \eqref{eq:M_general},\eqref{eq:M_P} can now be read off by collecting the coefficients of plane waves $te^{i(\mathbf{k\cdot x}\pm |\mathbf{k}|t)}$ from \eqref{eq:linearized},\eqref{eq:quadratic},\eqref{eq:G}:
\begin{align}
 M_2(k_\mu;P_\mu) =  \hat M_2(k_\mu;P_\mu)\,\delta^3(\mathbf{P - k}) \ ; \quad \hat M_2(k_\mu;P_\mu) = i\alpha\sign(k_t)\int_0^\infty \frac{dt}{t}\,e^{i(P_t - k_t)t} \ , \label{eq:M_integral}
\end{align}
where we relabeled the interaction time from $t'$ to $t$. Note that $\hat M_2(k_\mu;P_\mu)$ depends only on the energies $k_t,P_t$. As anticipated, the integral in \eqref{eq:M_integral} is divergent at $t=0$. We can apply dimensional regularization, multiplying the integrand by $t^\varepsilon$ with $\varepsilon>0$. Denoting $P_t - k_t \equiv E$ for brevity, the integral becomes:
\begin{align}
 \int_0^\infty dt\,t^{\varepsilon-1}e^{iEt} = \Gamma(\varepsilon)\left(\frac{i}{E}\right)^\varepsilon \underset{\varepsilon\rightarrow 0}{\longrightarrow} \ \frac{1}{\varepsilon} - \ln|E| + \frac{\pi i}{2}\sign(E) \ ,
\end{align}
which brings the evolution function \eqref{eq:M_integral} into the form:
\begin{align}
 \hat M_2(k_\mu;P_\mu) = i\alpha\sign(k_t)\left(\frac{1}{\varepsilon} - \ln|P_t - k_t| + \frac{\pi i}{2}\sign(P_t - k_t) \right) \ . \label{eq:M_P_divergent}
\end{align}
The divergent $1/\varepsilon$ term has no dependence on $P_\mu$, and thus on the incoming 4-momenta $p_\mu,q_\mu$. As a result, it will cancel whenever we integrate \eqref{eq:M_P_divergent} against an even combination $a(-|\mathbf{p}|,\mathbf{p}) = -a(|\mathbf{p}|,\mathbf{p})$ of incoming waves.

\section{Implementing the spacetime inversions} \label{sec:inversion}

In this section, we calculate explicitly the inversion kernels $\bbI_1(p_\mu,\tilde p_\mu)$ and $\bbI_2(P_\mu,\tilde P_\mu)$ from eqs. \eqref{eq:I_1},\eqref{eq:I_2}. Since these can be viewed as implementing inversions in flat spacetime, we will forget about de Sitter space in this section; in particular, we will raise/lower indices with the Minkowski metric $\eta_{\mu\nu}$. As we will see, both $\bbI_1$ and $\bbI_2$ can be expressed in terms of Bessel functions. Our main trick in the derivation will be to take spinor square roots of the 4-momenta. This will turn plane waves into Gaussians, and spacetime inversion into a Fourier transform between one Gaussian and another. 

\subsection{Inverting lightlike plane waves}

We begin with inversions \eqref{eq:I_1} of lightlike waves in Minkowski space. For this calculation, we introduce left-handed spinor indices $(\alpha,\beta,\dots)$ and right-handed ones $(\dot\alpha,\dot\beta,\dots)$, raised and lowered by the respective 2d Levi-Civita symbols:
\begin{align}
 \psi_\alpha = \epsilon_{\alpha\beta}\psi^\beta \ ; \quad \psi^\beta = \psi_\alpha\epsilon^{\alpha\beta} \ ; \quad \psi_{\dot\alpha} = \epsilon_{\dot\alpha\dot\beta}\psi^{\dot\beta} \ ; \quad \psi^{\dot\beta} = \psi_{\dot\alpha}\epsilon^{\dot\alpha\dot\beta} \ .
\end{align}
The spinor indices are related to vector indices $(\mu,\nu,\dots)$ via the 4d Pauli matrices $\sigma_\mu^{\alpha\dot\alpha}$, which satisfy: 
\begin{align}
 \sigma^\mu_{\alpha\dot\alpha}\sigma_\nu^{\alpha\dot\alpha} &= -2\delta^\mu_\nu \ ; \quad \sigma_\mu^{\alpha\dot\alpha}\sigma^\mu_{\beta\dot\beta} = -2\delta^\alpha_\beta \delta^{\dot\alpha}_{\dot\beta} \ ; \quad
 \sigma_{(\mu}^{\alpha\dot\alpha} \sigma_{\nu)\beta\dot\alpha} = -\eta_{\mu\nu}\delta^\alpha_\beta \ ; \quad \sigma_{(\mu}^{\alpha\dot\alpha} \sigma_{\nu)\alpha\dot\beta} = -\eta_{\mu\nu}\delta^{\dot\alpha}_{\dot\beta} \ .
\end{align}
For concreteness, we'll assume that the lightlike 4-momentum $\tilde p_\mu$ is future-pointing. As we will now see (and as already assumed implicitly in eqs. \eqref{eq:S_general},\eqref{eq:I_1}), it will transform under inversion into 4-momenta $p_\mu$ that are again lightlike and future-pointing. The case with past-pointing $p_\mu,\tilde p_\mu$ is completely analogous, and can be obtained by flipping the sign of $\sigma_t^{\alpha\dot\alpha}$, which is arbitrary anyway. We express $p_\mu$ and $\tilde p_\mu$ in terms of spinors as:
\begin{align}
 p_\mu = \sigma_\mu^{\alpha\dot\alpha}\lambda_\alpha\bar\lambda_{\dot\alpha} \ ; \quad \tilde p_\mu = \sigma_\mu^{\alpha\dot\alpha}\mu_\alpha\bar\mu_{\dot\alpha} \ ; \quad 
 p_\mu \tilde p^\mu = -2(\lambda_\alpha\mu^\alpha)(\bar\lambda_{\dot\alpha}\bar\mu^{\dot\alpha}) \ ,
\end{align}
where $\bar\lambda_{\dot\alpha},\bar\mu_{\dot\alpha}$ are the complex conjugates of $\lambda_\alpha,\mu_\alpha$. Note that the correspondence between the real, lightlike $p_\mu$ and the complex $\lambda_\alpha$ is not one-to-one: there is a residual $U(1)$ summetry of phase rotations $\lambda_\alpha\rightarrow e^{i\phi}\lambda_\alpha$ that preserve $p_\mu$. 

The spacetime position $x^\mu$ becomes a spinor matrix, with determinant and inverse given by:
\begin{align}
 x^{\alpha\dot\alpha} \equiv x^\mu\sigma_\mu^{\alpha\dot\alpha} \ ; \quad \det(x^{\alpha\dot\alpha}) = -x_\mu x^\mu \ ; \quad 
 (x^{-1})_{\dot\alpha\alpha} = -\frac{x_\mu\sigma^\mu_{\alpha\dot\alpha}}{x_\nu x^\nu} = -\tilde x_\mu\sigma^\mu_{\alpha\dot\alpha} \equiv -\tilde x_{\alpha\dot\alpha} \ .
\end{align}
Thus, spacetime inversion $x^\mu\rightarrow \tilde x^\mu$ is basically a matrix inversion of $x^{\alpha\dot\alpha}$. A lightlike plane wave can now be written as a Gaussian in $\lambda_\alpha,\bar\lambda_{\dot\alpha}$:
\begin{align}
 e^{ip_\mu x^\mu} = e^{ix^{\alpha\dot\alpha}\lambda_\alpha\bar\lambda_{\dot\alpha}} \ ,
\end{align}
and its inversion as a Fourier transform:
\begin{align}
 \frac{e^{i\tilde p_\mu x^\mu/(x_\nu x^\nu)}}{x_\nu x^\nu} = -\frac{e^{i\tilde x_{\alpha\dot\alpha}\mu^\alpha\bar\mu^{\dot\alpha}}}{\det(x^{\alpha\dot\alpha})} 
    = \frac{1}{\pi^2}\int d^4\lambda\,e^{ix^{\alpha\dot\alpha}\lambda_\alpha\bar\lambda_{\dot\alpha}}\,e^{i(\lambda_\alpha\mu^\alpha + \bar\lambda_{\dot\alpha}\bar\mu^{\dot\alpha})} \ , \label{eq:I_1_Fourier}
\end{align}
where $\int d^4\lambda$ is a $\bbC^2\cong \bbR^4$ integral over the real and imaginary parts of $\lambda_\alpha$. What remains is to express this $d^4\lambda$ integral as an integral over momenta $p_\mu$, and integrate out the residual $U(1)$ symmetry $\lambda_\alpha\rightarrow e^{i\phi}\lambda_\alpha$. We begin by expressing the Fourier phase $\lambda_\alpha\mu^\alpha + \bar\lambda_{\dot\alpha}\bar\mu^{\dot\alpha}$ in the integrand as:
\begin{align}
 2\Re(\lambda_\alpha\mu^\alpha) = 2\Re\left(\sqrt{-\frac{p_\mu \tilde p^\mu}{2}}\,e^{i\phi} \right) = \sqrt{-2p_\mu \tilde p^\mu}\,\cos\phi \ ,
\end{align}
where $\phi$ is the $U(1)$ phase of $\lambda_\alpha$, and we chose $\phi=0$ as the value for which $\lambda_\alpha\mu^\alpha$ is real. As for the integration measure $d^4\lambda$, it becomes:
\begin{align}
 d^4\lambda = \frac{d^3\mathbf{p}}{8|\mathbf{p}|}\,d\phi \ . \label{eq:d_lambda}
\end{align}
The numerical coefficient in \eqref{eq:d_lambda} is basically $\frac{1}{2}$ for each of the degrees of freedom in $\mathbf{p}$. For example, in spherical coordinates, the magnitude $|\lambda|\equiv \sqrt{|\lambda_0|^2 + |\lambda_1|^2}$ is $\sqrt{|\mathbf{p}|}$, so differentiating it gives a factor of $\frac{1}{2}$; another factor of $\frac{1}{2}$ then comes from each of the two angles that determine the direction of $\mathbf{p}$, which are halved in the spinor representation. 

Putting everything together, we obtain the inversion kernel \eqref{eq:I_1} as:
\begin{align}
 \bbI_1(p_\mu,\tilde p_\mu) = \frac{1}{4\pi^2}\int_0^{2\pi} d\phi\, e^{i\sqrt{-2p_\mu \tilde p^\mu}\,\cos\phi} = \frac{1}{2\pi}J_0\!\left(\!\sqrt{-2p_\mu \tilde p^\mu} \right) \ ,
\end{align}
where $J_n$ is the Bessel function of the first kind. We can now write down explicitly the \emph{linearized} approximation to the horizon$\rightarrow$horizon evolution \eqref{eq:S_general} as:
\begin{align}
  a_H(k_\mu) = \frac{1}{2\pi}\int_{\substack{\tilde k^2\,=\,0 \\ k\cdot\tilde k\,<\,0}} \frac{d^3\mathbf{\tilde k}}{2|\mathbf{\tilde k}|}\,J_0\!\left(\!\sqrt{-2k_\mu \tilde k^\mu} \right) \tilde a_{\tilde H}(\tilde k_\mu) + O(\tilde a_{\tilde H}^2) \ .
\end{align}
This was presented in \cite{David:2019mos} in the spinor language, i.e. without performing the $\phi$ integral.

\subsection{Inverting timelike plane waves}

To evaluate the interacting term \eqref{eq:S2_integrated} in the horizon$\rightarrow$horizon evolution \eqref{eq:S_general}, we will need to also know the kernel $\bbI_2(P_\mu,\tilde P_\mu)$ for inversions with non-lightlike 4-momentum. As in the lightlike case, we will see that the causal nature of the 4-momentum is preserved: a plane wave with spacelike 4-momentum $\tilde P_\mu$ inverts into a superposition of spacelike 4-momenta $P_\mu$, while a wave with timelike $\tilde P_\mu$ inverts into a superposition of timelike $P_\mu$ with the same time-orientation.  We begin with the timelike case, which is a bit easier. Again, it will be sufficient to consider future-pointing $P_\mu,\tilde P_\mu$: the past-pointing case follows trivially.

Let's now express a future-pointing timelike 4-momentum $P_\mu$ in terms of spinors. This can be done by temporarily undoing eq. \eqref{eq:P}, and expressing the timelike $P_\mu$ as a sum of two lightlike vectors. Each of these can be composed as before out of Weyl spinors. Thus, we introduce two left-handed Weyl spinors $\lambda_\alpha,\lambda'_\alpha$ along with their right-handed counterparts $\bar\lambda_{\dot\alpha},\bar\lambda'_{\dot\alpha}$, and write:
\begin{align}
 P_\mu = \sigma_\mu^{\alpha\dot\alpha}(\lambda_\alpha\bar\lambda_{\dot\alpha} + \lambda'_\alpha\bar\lambda'_{\dot\alpha}) \ . \label{eq:P_spinors}
\end{align}
This is just the famous construction $P_\mu = \bar\Psi\gamma_\mu\Psi$ of the 4-current (or 4-momentum) of a massive electron in terms of a Dirac spinor $\Psi = (\lambda_\alpha,\bar\lambda'_{\dot\alpha})$. However, in our context, it will be more convenient to think of the left-handed spinors $\lambda_\alpha,\lambda'_\alpha$ on an equal footing, without combining one of them with the complex conjugate of the other. For $\tilde P_\mu$, we define similarly:
\begin{align}
 \tilde P_\mu = \sigma_\mu^{\alpha\dot\alpha}(\mu_\alpha\bar\mu_{\dot\alpha} + \mu'_\alpha\bar\mu'_{\dot\alpha}) \ . \label{eq:tilde_P}
\end{align}
The inversion \eqref{eq:I_2} can now be expressed as a product of two lightlike inversions, just like in the second line of \eqref{eq:S2_raw}: the first inverting $\lambda_\alpha\bar\lambda_{\dot\alpha}$ into $\mu_\alpha\dot\mu_{\dot\alpha}$, and the second inverting $\lambda'_\alpha\bar\lambda'_{\dot\alpha}$ into $\mu'_\alpha\dot\mu'_{\dot\alpha}$. As we saw in \eqref{eq:I_1_Fourier}, each of these lightlike inversions is a Fourier transform of the relevant spinor varialbes. Thus, we arrive at:
\begin{align}
 \frac{e^{i\tilde P_\mu x^\mu/(x_\nu x^\nu)}}{(x_\nu x^\nu)^2} = \frac{1}{\pi^4}\int d^4\lambda\,d^4\lambda'\,e^{iP_\mu x^\mu} 
   e^{i(\lambda_\alpha\mu^\alpha + \lambda'_\alpha\mu'^\alpha + \bar\lambda_{\dot\alpha}\bar\mu^{\dot\alpha} + \bar\lambda'_{\dot\alpha}\bar\mu'^{\dot\alpha})} \ . \label{eq:I_2_Fourier}
\end{align}
At this point, we switch back to treating the timelike vector $P_\mu$ as a whole. As in the lightlike case, the construction \eqref{eq:P_spinors} of $P_\mu$ (which has 4 real components) out of the spinors $\lambda_\alpha,\lambda'_\alpha$ (which have 8 real components) comes with a residual symmetry, this time a 4-dimensional one. Indeed, eq. \eqref{eq:P_spinors} is invariant under transformations:
\begin{align}
 \lambda_\alpha \rightarrow g_\alpha{}^\beta\lambda_\beta \ ; \quad \lambda'_\alpha \rightarrow g_\alpha{}^\beta\lambda'_\beta \ , \label{eq:residual_symmetry}
\end{align}
where $g$ belongs to a $U(2)=SU(2)\times U(1)$ subgroup of $SL(2,\bbC)$. Here, the $SU(2)$ describes 3d spatial rotations around the direction of $P_\mu$, while the $U(1)$ is an overall phase rotation of the spinors. Our task now is to convert the spinor integral $d^4\lambda\,d^4\lambda'$ into a 4-momentum integral $d^4P$, and integrate out this residual $U(2)$ symmetry. 

We begin by choosing reference values for the integration variables $\lambda_\alpha,\lambda'_\alpha$:
\begin{align}
 \hat\lambda_\alpha &= \sqrt{\frac{m}{2}}\begin{pmatrix} 1 \\ 0 \end{pmatrix} \ ; \quad \hat\lambda'_\alpha = \sqrt{\frac{m}{2}}\begin{pmatrix} 0 \\ 1 \end{pmatrix} \ . \label{eq:lambda_hat}
\end{align}
These describe a 4-momentum $P^\mu = (m,0,0,0)$ of length $m$, oriented along the time axis. Now, an arbitrary point in $(\lambda_\alpha,\lambda'_\alpha)$ space can be parameterized analogously to \eqref{eq:residual_symmetry}:
\begin{align}
 \lambda_\alpha = G_\alpha{}^\beta \hat\lambda_\beta \ ; \quad \lambda'_\alpha = G_\alpha{}^\beta \hat\lambda'_\beta \ ,
\end{align}
where $G_\alpha{}^\beta\in GL(2,\bbC)$ is now an arbitrary complex matrix. Indeed, the components of $\lambda_\alpha$ and $\lambda'_\alpha$ are just the first and second columns of $G_\alpha{}^\beta$, up to the factors of $\sqrt{m/2}$ in \eqref{eq:lambda_hat}. It is convenient to impose a group-invariant metric on $GL(2,\bbC)$:
\begin{align}
 (d^2s)_{GL(2,\bbC)} = \frac{1}{2}\Re\!\left[\tr(G^{-1}dG)^2 \right] \ . \label{eq:dG}
\end{align}
At the identity $G_\alpha{}^\beta = \delta_\alpha^\beta$, this metric assigns unit norm to the standard $GL(2,\bbC)$ generators: 
\begin{align}
 &G_\alpha{}^\beta = \delta_\alpha^\beta \ , \quad dG_\alpha{}^\beta = \begin{pmatrix} (a+b) + i(c+d) & (e-f) + i(g-h) \\ (e+f) + i(g+h) & (a-b) + i(c-d) \end{pmatrix} \\
 &\ \Longrightarrow \ (d^2s)_{GL(2,\bbC)} = a^2 + b^2 + e^2 + h^2 - c^2 - d^2 - f^2 -  g^2 \ .
\end{align}
Our integration measure at $G_\alpha{}^\beta=\delta_\alpha^\beta$ can now be rewritten as:
\begin{align}
 d^4\lambda\,d^4\lambda' = m^4 d^8G \ ,
\end{align}
where $d^8G$ is the measure associated with the metric \eqref{eq:dG}, i.e. $1/2^4$ times the trivial measure over the real and imaginary parts of the matrix elements $G_\alpha{}^\beta$. Now, this $d^8G$ decomposes into two orthogonal 4d pieces. First, we have the space of unitary matrices $g\in U(2)$, which describe the residual symmetry \eqref{eq:lambda_hat} that preserves the reference vector $P^\mu=(m,0,0,0)$. Second, we have the orthogonal complement of $U(2)$, which changes $P^\mu$ via boosts and rescalings. Again, in the spinor representation, the scale factor and the boost angles are related to those in the vector representation by a square root and by factors of $\frac{1}{2}$, respectively. Taking these into account, we write the measure as:
\begin{align}
 d^8G = \frac{d^4P}{(2m)^4}\,d^4g \ ,
\end{align}
where the factor of $1/2^4$ essentially consists of $\frac{1}{2}$ for each component of $P^\mu$. Altogether, we conclude that the measure in \eqref{eq:I_2_Fourier} decomposes as:
\begin{align}
 d^4\lambda\,d^4\lambda' = \frac{1}{2^4}\,d^4P\,d^4g \ . \label{eq:measure_decomposition_timelike}
\end{align}
Thanks to our group-invariant definition \eqref{eq:dG} of the group metric, we are now free to use eq. \eqref{eq:measure_decomposition_timelike} everywhere, not only in the infinitesimal neighborhood of $G_\alpha{}^\beta = \delta_\alpha^\beta$, i.e. of the reference spinors \eqref{eq:lambda_hat}. 

Let us now integrate out the residual symmetry $g$. We again denote the timelike length of $P^\mu$ as $m$. Similarly, let $\tilde m$ denote the timelike length of $\tilde P^\mu$, and let $\chi$ denote the boost angle between $P^\mu$ and $\tilde P^\mu$:
\begin{align}
 P_\mu P^\mu = -m^2 \ ; \quad \tilde P_\mu\tilde P^\mu = -\tilde m^2 \ ; \quad P_\mu\tilde P^\mu = -m\tilde m\cosh\chi \ .
\end{align}
We choose a Lorentz frame such that $P^\mu$ again points along the time axis, $P^\mu = (m,0,0,0)$, while $\tilde P^\mu$ lies in the $tz$ plane, $\tilde P^\mu = \tilde m(\cosh\chi,0,0,\sinh\chi)$. This value of $\tilde P^\mu$ can be composed from the spinors:
\begin{align}
 \mu_\alpha &= \sqrt{\frac{\tilde m}{2}}\begin{pmatrix} e^{\chi/2} \\ 0 \end{pmatrix} \ ; \quad \mu'_\alpha = \sqrt{\frac{\tilde m}{2}}\begin{pmatrix} 0 \\ e^{-\chi/2} \end{pmatrix} \ . \label{eq:mu}
\end{align} 
Now, the spinors $\lambda_\alpha,\lambda'_\alpha$ that make up $P_\mu$ will be given by \eqref{eq:lambda_hat}, multiplied by $g\in U(2)$. This can be parameterized as:
\begin{align}
 g_\alpha{}^\beta &= e^{i\phi}\begin{pmatrix} \cos\theta\,e^{i\beta} & -\sin\theta\,e^{-i\gamma} \\ \sin\theta\,e^{i\gamma} & \cos\theta\,e^{-i\beta} \end{pmatrix} \ ; \label{eq:g_timelike} \\
 \lambda_\alpha &= g_\alpha{}^\beta\hat\lambda_\beta = \sqrt{\frac{m}{2}}\,e^{i\phi}\begin{pmatrix} \cos\theta\,e^{i\beta} \\ \sin\theta\,e^{i\gamma} \end{pmatrix} \ ; \quad 
 \lambda'_\alpha = g_\alpha{}^\beta\hat\lambda'_\beta = \sqrt{\frac{m}{2}}\,e^{i\phi}\begin{pmatrix} -\sin\theta\,e^{-i\gamma} \\ \cos\theta\,e^{-i\beta} \end{pmatrix} \ , \label{eq:lambda_g_timelike}
\end{align}
where the ranges are $\beta,\gamma\in(0,2\pi)$, $\theta\in(0,\frac{\pi}{2})$ and $\phi\in(0,\pi)$. The measure induced by the metric \eqref{eq:dG} on the $U(2)$ matrices \eqref{eq:g_timelike} is:
\begin{align}
 d^4g = d\theta(\cos\theta d\beta)(\sin\theta d\gamma)d\phi = \sin\theta\,d(\sin\theta)d\beta d\gamma d\phi \ . \label{eq:measure_g_timelike}
\end{align}
The Fourier phase in the integrand of \eqref{eq:I_2_Fourier} now takes the form:
\begin{align}
 \begin{split}
   2\Re(\lambda_\alpha\mu^\alpha + \lambda'_\alpha\mu'^\alpha) &= \sqrt{m\tilde m}\sin\theta\Re\!\left(e^{\chi/2}e^{i(\phi+\gamma)} + e^{-\chi/2} e^{i(\phi-\gamma)}\right) \\
   &= \sin\theta(A_+\cos(\phi+\gamma) + A_-\cos(\phi-\gamma)) \ , 
 \end{split} \label{eq:Fourier_phase_timelike}
\end{align}
where we denoted:
\begin{align}
 \begin{split}
   &A_\pm(P_\mu,\tilde P_\mu) \equiv \sqrt{m\tilde m}\,e^{\pm\chi/2} \\
     &\qquad = \frac{1}{\sqrt{2}}\left( \sqrt{-P_\mu\tilde P^\mu + \sqrt{(P_\mu P^\mu)(\tilde P_\nu\tilde P^\nu)}} \pm \sqrt{-P_\mu\tilde P^\mu - \sqrt{(P_\mu P^\mu)(\tilde P_\nu\tilde P^\nu)}} \right) \ . 
 \end{split} \label{eq:A_timelike}
\end{align}
Note that the phase \eqref{eq:Fourier_phase_timelike} doesn't depend on $\beta$, and has a clean, factorized dependence on $\theta$ and $\phi\pm\gamma$. Plugging eqs. \eqref{eq:measure_decomposition_timelike} and \eqref{eq:measure_g_timelike}-\eqref{eq:Fourier_phase_timelike} back into \eqref{eq:I_2_Fourier}, we obtain the inversion kernel \eqref{eq:I_2} in the form:
\begin{align}
 \bbI_2(P_\mu,\tilde P_\mu) = \frac{1}{32\pi^4} \int_0^1 \sin\theta\,d(\sin\theta) \int_0^{2\pi}d\beta \int_0^{2\pi}d\gamma\int_0^{2\pi}d\phi\,e^{i\sin\theta(A_+\cos(\phi+\gamma) + A_-\cos(\phi-\gamma))} \ ,
\end{align}
where we doubled the range of $\phi$ for uniformity with the other angles, and compensated with a factor of $\frac{1}{2}$. Let us now switch variables from $\phi,\gamma$ to $\phi_{\pm} \equiv \phi\pm\gamma$, while maintaining an integration range of $(2\pi)^2$ over the two angles. Performing the trivial $\beta$ integral and renaming $\sin\theta\equiv \xi$, we arrive at:
\begin{align}
 \bbI_2(P_\mu,\tilde P_\mu) = \frac{1}{16\pi^3}\int_0^1 \xi d\xi \int_0^{2\pi}d\phi_+ \int_0^{2\pi}d\phi_-\,e^{i\xi(A_+\cos\phi_+ + A_-\cos\phi_-)} \ .
\end{align}
The $\phi_\pm$ integrals factorize, each yielding a Bessel function $J_0$. We thus get:
\begin{align}
 \begin{split}
   \bbI_2(P_\mu,\tilde P_\mu) &= \frac{1}{4\pi}\int_0^1 \xi d\xi\,J_0(A_+\xi) J_0(A_-\xi) \\
     &= \frac{1}{4\pi}\cdot\left.\frac{A_+ \xi\,J_1(A_+\xi) J_0(A_-\xi) - A_- \xi\,J_1(A_-\xi) J_0(A_+\xi)}{A_+^2 - A_-^2} \right|_{\xi = 0}^1 \\
     &= \frac{A_+ J_1(A_+) J_0(A_-) - A_- J_1(A_-) J_0(A_+)}{4\pi(A_+^2 - A_-^2)} \ ,
 \end{split} \label{eq:I_2_timelike}
\end{align}
where $A_\pm$ are the functions of $P_\mu,\tilde P_\mu$ given by \eqref{eq:A_timelike}.

\subsection{Inverting spacelike plane waves}

We now turn to the case of spacelike $\tilde P_\mu$, which will invert into a superposition of spacelike $P_\mu$. We follow the same strategy as with the timelike case, writing $P_\mu,\tilde P_\mu$ as sums of two lightlike vectors, this time with opposite time orientations:
\begin{align}
 P_\mu = \sigma_\mu^{\alpha\dot\alpha}(\lambda_\alpha\bar\lambda_{\dot\alpha} - \lambda'_\alpha\bar\lambda'_{\dot\alpha}) \ ; \label{eq:P_spacelike} \\
 \tilde P_\mu = \sigma_\mu^{\alpha\dot\alpha}(\mu_\alpha\bar\mu_{\dot\alpha} - \mu'_\alpha\bar\mu'_{\dot\alpha}) \ . \label{eq:tilde_P_spacelike}
\end{align}
In Dirac-spinor notation, this construction is just $P_\mu = \bar\Psi\gamma_\mu\gamma_5\Psi$, which produces e.g. the angular momentum vector of a Dirac electron.

Eq. \eqref{eq:I_2_Fourier}, which expresses the inversion as a spinor Fourier transform, stays unchanged. The residual symmetry of transformations $g_\alpha{}^\beta$ that preserve $P_\mu$ is now $SL(2,\bbR)\times U(1)$, where $SL(2,\bbR)$ is the Lorentz group in the 3d hyperplane perpendicular to $P_\mu$, and $U(1)$ is again an overall phase rotation. The decomposition \eqref{eq:measure_decomposition_timelike} of the spinor measure into $d^4P$ and $d^4g$ takes the same form as before. 

However, now comes a complication that did not appear in the timelike case. A priori, the spinor Fourier transform \eqref{eq:Fourier_phase_timelike} turns any 4-momentum of the form \eqref{eq:tilde_P_spacelike} into a superposition of all possible 4-momenta of the form \eqref{eq:P_spacelike}, i.e. a spacelike $\tilde P_\mu$ becomes a superposition of all possible spacelike $P_\mu$. However, there are three distinct ways in which such vectors can be related:
\begin{enumerate}[label=\Roman*.]
	\item $P_\mu,\tilde P_\mu$ lie in the same quadrant of a timelike plane, separated by a boost angle $\chi$.
	\item $P_\mu,\tilde P_\mu$ lie in opposite quadrants of a timelike plane, such that $P_\mu$ and $-\tilde P_\mu$ are separated by a boost angle $\chi$.
	\item $P_\mu,\tilde P_\mu$ lie in a spacelike plane, separated by an angle $\chi$.
\end{enumerate} 
These three possibilities describe three regions of the $d^4P$ integral in \eqref{eq:I_2}. We will now calculate the kernel $\bbI_2(P_\mu,\tilde P_\mu)$ in each of these regions separately.
\subsubsection*{Region I}
Here, we have:
\begin{align}
 P_\mu P^\mu = m^2 \ ; \quad \tilde P_\mu\tilde P^\mu = \tilde m^2 \ ; \quad P_\mu \tilde P^\mu = m\tilde m\cosh\chi \ ,
\end{align}
where $m,\tilde m$ are now the \emph{spacelike} lengths of $P_\mu,\tilde P_\mu$ respectively. In an adapted Lorentz frame, we can set $P^\mu = (0,0,0,m)$ and $\tilde P^\mu = \tilde m(\sinh\chi,0,0,\cosh\chi)$, and represent these vectors with the same spinors \eqref{eq:lambda_hat},\eqref{eq:mu} that we used in the timelike case. Despite its group structure, the $SL(2,\bbR)$ symmetry that preserves the direction of $P^\mu$ (in this case, the $z$ axis) is not represented by real matrices; instead, it takes the same form as the $SU(2)$ in \eqref{eq:g_timelike}, but with the angle $\theta$ turned hyperbolic, with range $\theta\in(0,\infty)$. Overall, the residual  $SL(2,\bbR)\times U(1)$ symmetry and its action on the reference spinors $\hat\lambda_\alpha,\hat\lambda'_\alpha$ take the form:
\begin{align}
 g_\alpha{}^\beta &= e^{i\phi}\begin{pmatrix} \cosh\theta\,e^{i\beta} & \sinh\theta\,e^{-i\gamma} \\ \sinh\theta\,e^{i\gamma} & \cosh\theta\,e^{-i\beta} \end{pmatrix} \ ; \label{eq:g_spacelike_I} \\
 \lambda_\alpha &= g_\alpha{}^\beta\hat\lambda_\beta = \sqrt{\frac{m}{2}}\,e^{i\phi}\begin{pmatrix} \cosh\theta\,e^{i\beta} \\ \sinh\theta\,e^{i\gamma} \end{pmatrix} \ ; \quad 
 \lambda'_\alpha = g_\alpha{}^\beta\hat\lambda'_\beta = \sqrt{\frac{m}{2}}\,e^{i\phi}\begin{pmatrix} \sinh\theta\,e^{-i\gamma} \\ \cosh\theta\,e^{-i\beta} \end{pmatrix} \ . \label{eq:lambda_g_spacelike_I}
\end{align}
The measure $d^4g$ takes the same form as in \eqref{eq:measure_g_timelike}, but with $\sinh\theta$ instead of $\sin\theta$:
\begin{align}
 d^4g = d\theta(\cosh\theta d\beta)(\sinh\theta d\gamma)d\phi = \sinh\theta\,d(\sinh\theta)d\beta d\gamma d\phi \ . \label{eq:measure_g_spacelike_I}
\end{align}
The Fourier phase in \eqref{eq:I_2_Fourier} becomes:
\begin{align}
  2\Re(\lambda_\alpha\mu^\alpha + \lambda'_\alpha\mu'^\alpha) = \sinh\theta(A_+\cos(\phi+\gamma) - A_-\cos(\phi-\gamma)) \ , \label{eq:Fourier_phase_spacelike_I}
\end{align}
where $A_\pm$ are defined similarly to \eqref{eq:A_timelike}, but without minus signs in front of $P_\mu\tilde P^\mu$:
\begin{align}
 \begin{split} 
    &A_\pm(P_\mu,\tilde P_\mu) \equiv \sqrt{m\tilde m}\,e^{\pm\chi/2} \\
    &\qquad = \frac{1}{\sqrt{2}}\left( \sqrt{P_\mu\tilde P^\mu + \sqrt{(P_\mu P^\mu)(\tilde P_\nu\tilde P^\nu)}} \pm \sqrt{P_\mu\tilde P^\mu - \sqrt{(P_\mu P^\mu)(\tilde P_\nu\tilde P^\nu)}} \right) \ . 
 \end{split} \label{eq:A_spacelike_I}
\end{align}
Putting everything together, we obtain the same integral as in \eqref{eq:I_2_timelike}, but with $\xi \equiv \sinh\theta\in(0,\infty)$ instead of $\xi \equiv \sin\theta\in(0,1)$:
\begin{align}
  \bbI_2(P_\mu,\tilde P_\mu) = \frac{1}{4\pi}\int_0^\infty \xi d\xi\,J_0(A_+\xi) J_0(A_-\xi) = \frac{1}{4\pi A_+}\delta(A_+ - A_-) \ . \label{eq:I_2_spacelike_delta}
\end{align}
Here, we used an orthogonality property of the Bessel functions, which can be derived e.g. as the $n=0,\varepsilon\rightarrow 0$ limit of Weber's second integral:
\begin{align}
 \int_0^\infty \xi d\xi\,J_n(A_+\xi) J_n(A_-\xi)\,e^{-\varepsilon^2\xi^2} = \frac{1}{2\varepsilon^2}\, e^{-\frac{A_+^2 + A_-^2}{4\varepsilon^2}} I_n\!\left(\frac{A_+ A_-}{2\varepsilon^2}\right) \ ,
\end{align}
where $I_n$ is the modified Bessel function of the first kind. 

In our context, the delta function on the RHS of \eqref{eq:I_2_spacelike_delta} can be neglected. Indeed, $A_+ - A_-$ vanishes at the \emph{edge} of the momentum-space region we're considering, where the boost angle $\chi$ goes to zero, and the plane of $(P_\mu,\tilde P_\mu)$ becomes lightlike. Near this edge, $A_+ - A_-$ depends on the components of $P_\mu$ (e.g. on $P_t$) as a square root $\sqrt{f(P_t)}$, where $f$ and its derivative don't generally vanish together. The delta function \eqref{eq:I_2_spacelike_delta} thus enters into the $d^4P$ integral as $\sim \delta\big(\sqrt{f(P_t)}\big)dP_t$, which vanishes. We conclude that Region I does not contribute to the inversion \eqref{eq:I_2}:
\begin{align}
 \bbI_2(P_\mu,\tilde P_\mu) = 0 \ . \label{eq:I_2_spacelike_I}
\end{align} 
\subsubsection*{Region II}
Here, we have:
\begin{align}
 P_\mu P^\mu = m^2 \ ; \quad \tilde P_\mu\tilde P^\mu = \tilde m^2 \ ; \quad P_\mu \tilde P^\mu = -m\tilde m\cosh\chi \ .
\end{align}
We can keep our parameterization of $P_\mu$, $\hat\lambda_\alpha$, $\hat\lambda'_\alpha$ and $g_\alpha{}^\beta$ from Region I. However, we now set $\tilde P^\mu$ to minus its previous value, i.e. $\tilde P^\mu = -\tilde m(\sinh\chi,0,0,\cosh\chi)$. This vector can be constructed by interchanging the two spinors $\mu_\alpha,\mu'_\alpha$ in \eqref{eq:mu}:
\begin{align}
 \mu_\alpha &= \sqrt{\frac{\tilde m}{2}}\begin{pmatrix} 0 \\ e^{-\chi/2} \end{pmatrix} \ ; \quad \mu'_\alpha = \sqrt{\frac{\tilde m}{2}}\begin{pmatrix} e^{\chi/2} \\ 0 \end{pmatrix} \ . \label{eq:mu_II}
\end{align}
The Fourier phase in \eqref{eq:I_2_Fourier} now reads:
\begin{align}
 2\Re(\lambda_\alpha\mu^\alpha + \lambda'_\alpha\mu'^\alpha) = \cosh\theta(A_+\cos(\phi-\beta) - A_-\cos(\phi+\beta)) \ , \label{eq:Fourier_phase_spacelike_II}
\end{align}
where $A_\pm$ are given by the same expression \eqref{eq:A_timelike} as in the timelike case. Because of the appearance of $\cosh\theta$ in \eqref{eq:Fourier_phase_spacelike_II}, it is convenient to rewrite the measure \eqref{eq:measure_g_spacelike_I} as:
\begin{align}
 d^4g = \cosh\theta\,d(\cosh\theta)d\beta d\gamma d\phi \ . \label{eq:measure_g_spacelike_II}
\end{align}
We end up with the same kind of integral as in \eqref{eq:I_2_timelike} and \eqref{eq:I_2_spacelike_delta}, but now with $\xi\equiv\cosh\theta\in(1,\infty)$. Due to the vanishing \eqref{eq:I_2_spacelike_delta},\eqref{eq:I_2_spacelike_I} of the integral from 0 to $\infty$, this is just minus the integral from \eqref{eq:I_2_timelike}:
\begin{align} 
 \begin{split}
   \bbI_2(P_\mu,\tilde P_\mu) &= \frac{1}{4\pi}\int_1^\infty \xi d\xi\,J_0(A_+\xi) J_0(A_-\xi) = -\frac{1}{4\pi}\int_0^1 \xi d\xi\,J_0(A_+\xi) J_0(A_-\xi) \\
     &= -\frac{A_+ J_1(A_+) J_0(A_-) - A_- J_1(A_-) J_0(A_+)}{4\pi(A_+^2 - A_-^2)} \ .
 \end{split}
\end{align}
\subsubsection*{Region III}
\begin{figure}%
	\centering%
	\includegraphics[scale=.8]{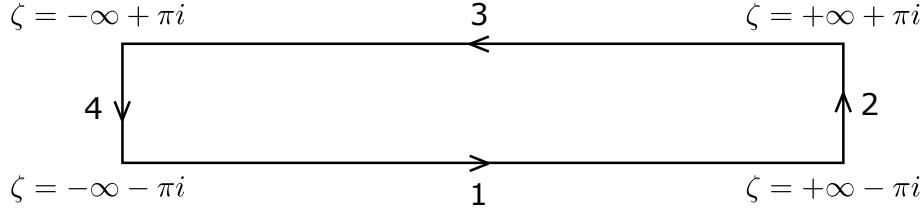} \\
	\caption{The integration contour $\del\Omega$ in the complex $\zeta$ plane from eq. \eqref{eq:zeta_contour}.}
	\label{fig:zeta} 
\end{figure}

The final region -- spacelike 4-momenta separated by a spatial rotation angle -- is the trickiest. Here, we have:
\begin{align}
P_\mu P^\mu = m^2 \ ; \quad \tilde P_\mu\tilde P^\mu = \tilde m^2 \ ; \quad P_\mu \tilde P^\mu = m\tilde m\cos\chi \ . 
\end{align}
This time, it will be convenient to fix our representative vectors as $P^\mu = (0,0,m,0)$ and $\tilde P^\mu = \tilde m(0,-\sin\chi,\cos\chi,0)$, where we take $\chi\in(0,\pi)$. These 4-momenta can be constructed from the spinors:
\begin{align}
  \hat\lambda_\alpha &= \frac{\sqrt{m}}{2}\begin{pmatrix} 1 \\ -i \end{pmatrix} \ ; \quad \hat\lambda'_\alpha = \frac{\sqrt{m}}{2}\begin{pmatrix} 1 \\ i \end{pmatrix} \ ; \\
  \mu_\alpha &= \frac{\sqrt{\tilde m}}{2}\begin{pmatrix} e^{i\chi/2} \\ -ie^{-i\chi/2} \end{pmatrix} \ ; \quad \mu'_\alpha = \frac{\sqrt{\tilde m}}{2}\begin{pmatrix} e^{i\chi/2} \\ ie^{-i\chi/2} \end{pmatrix} \ .
\end{align}
With $P^\mu$ now chosen along the $y$ axis, the $SL(2,\bbR)$ symmetry now literally consists of real $2\times 2$ matrices with unit determinant. These can be parameterized similarly to \eqref{eq:g_timelike},\eqref{eq:g_spacelike_I}, but at the cost of splitting $SL(2,\bbR)$ into different domains, distinguished by the signs of the different matrix elements:
\begin{align}
 \pm\!\begin{pmatrix} \eta\cosh\theta\,e^\beta & \sinh\theta\,e^{-\gamma} \\ \sinh\theta\,e^\gamma & \eta\cosh\theta\,e^{-\beta} \end{pmatrix} , \ 
 \pm\!\begin{pmatrix} \eta\cos\theta\,e^\beta & -\sin\theta\,e^{-\gamma} \\ \sin\theta\,e^\gamma & \eta\cos\theta\,e^{-\beta} \end{pmatrix} , \ 
 \pm\!\begin{pmatrix} \eta\sinh\theta\,e^\beta & -\cosh\theta\,e^{-\gamma} \\ \cosh\theta\,e^\gamma & -\eta\sinh\theta\,e^{-\beta} \end{pmatrix} \ .
\end{align}
We thus have 3 kinds of matrices, which will make up 3 integration domains. Within each one, there is a further choice of sign $\eta=\pm 1$ for the diagonal elements, and an additional overall sign on the entire matrix. $\eta$ will end up decoupling from the inversion integral \eqref{eq:I_2_Fourier}, so that the sum over its values will simply show up as a factor of 2. As for the matrix's overall sign, we will absorb it into the $U(1)$ phase freedom. Overall, we parameterize the residual $SL(2,\bbR)\times U(1)$ symmetry as:
\begin{align}
 \begin{split}
   \text{Domain 1:}\qquad g_\alpha{}^\beta &= e^{i\phi} \begin{pmatrix} \eta\cosh\theta\,e^\beta & \sinh\theta\,e^{-\gamma} \\ \sinh\theta\,e^\gamma & \eta\cosh\theta\,e^{-\beta} \end{pmatrix} \ ; \\
   \text{Domain 2:}\qquad g_\alpha{}^\beta &= ie^{i\phi} \begin{pmatrix} \eta\cos\theta\,e^\beta & -\sin\theta\,e^{-\gamma} \\ \sin\theta\,e^\gamma & \eta\cos\theta\,e^{-\beta} \end{pmatrix} \ ; \\
   \text{Domain 3:}\qquad g_\alpha{}^\beta &= ie^{i\phi} \begin{pmatrix} \eta\sinh\theta\,e^\beta & -\cosh\theta\,e^{-\gamma} \\ \cosh\theta\,e^\gamma & -\eta\sinh\theta\,e^{-\beta} \end{pmatrix} \ ,
 \end{split} \label{eq:g_spacelike_III}
\end{align}
where the factor of $i$ in Domains 2,3 is for later convenience. The parameter ranges in \eqref{eq:g_spacelike_III} are:
\begin{align}
 \eta=\pm 1 \ ; \quad \beta,\gamma\in(-\infty,\infty) \ ; \quad \phi\in(-\pi,\pi) \ ; \quad \theta \in \left\{\begin{array}{cl}
   	(0,\infty) & \quad \text{Domains 1,3} \\  
 	(0,\frac{\pi}{2}) & \quad \text{Domain 2}
 \end{array} \right. \ .
\end{align}
Recall that in the previous momentum regions, it was a good idea to switch variables from $\phi$ and $\gamma$ to $\phi\pm\gamma$, and from $\theta$ to its (ordinary or hyperbolic) sine. In the present case, we similarly define new variables as: 
\begin{align}
 \zeta \equiv \gamma + i\phi \ ; \quad \xi = \left\{\begin{array}{cl}
    \sinh\theta & \quad \text{Domain 1} \\  
    i\sin\theta & \quad \text{Domain 2} \\
    i\cosh\theta & \quad \text{Domain 3}
 \end{array} \right. \ , \label{eq:xi}
 \end{align}
where instead of $\phi\pm\gamma$ we now have $\zeta$ and its complex conjugate, and $\xi$ now ranges over the positive real \emph{and} imaginary axes. In terms of these variables, the $SL(2,\bbR)\times U(1)$ measure $d^4g$ (up to sign, which will be fixed by contour orientation) reads:
\begin{align}
d^4g = \frac{i}{2}\xi d\xi d\beta d\zeta d\bar\zeta \ ,
\end{align}
while the Fourier phase in \eqref{eq:I_2_Fourier} takes the compact form:
\begin{align}
  2\Re(\lambda_\alpha\mu^\alpha + \lambda'_\alpha\mu'^\alpha) = 2\Re(A\xi\cosh\zeta) \ .
\end{align}
Here, $A$ is a complex function of the momenta $P_\mu,\tilde P_\mu$:
\begin{align}
 \begin{split} 
    &A(P_\mu,\tilde P_\mu) = \sqrt{m\tilde m}\,e^{i\chi/2} \\
    &\qquad = \frac{1}{\sqrt{2}}\left( \sqrt{\sqrt{(P_\mu P^\mu)(\tilde P_\nu\tilde P^\nu)} + P_\mu\tilde P^\mu} + i\,\sqrt{\sqrt{(P_\mu P^\mu)(\tilde P_\nu\tilde P^\nu)} - P_\mu\tilde P^\mu} \right) \ , 
 \end{split} \label{eq:A_spacelike_III}
\end{align}
and we've arranged things such that $\Im(A\xi)>0$. Plugging everything into \eqref{eq:I_2_Fourier}, the inversion kernel takes the form:
\begin{align}
 \bbI_2(P_\mu,\tilde P_\mu) = \frac{i}{16\pi^4}\left( \int_0^\infty + \int_{i\infty}^0\right) \xi d\xi \int_{-\infty}^\infty d\beta \int_\Omega d\zeta d\bar\zeta\,e^{2i\Re(A\xi\cosh\zeta)} \ , \label{eq:I_big_integral}
\end{align}
where $\Omega$ is the strip $\Im\zeta\in(-\pi,\pi)$ in the complex $\zeta$ plane. As in the previous momentum regions, the integrand does not depend on $\beta$ (or on $\eta$, which we already summed over). In this case, though, the range of $\beta$ is infinite, making the integral diverge. As we will see, this divergence will cancel against a zero in the $\xi$ integral. For now, let us focus on the complex integral over $(\zeta,\bar\zeta)$. In the previous regions, the analogues $\phi\pm\gamma$ of $(\zeta,\bar\zeta)$ were independent real variables, and the integral over them factorized to give a product of Bessel functions. Our present situation is similar, but since $(\zeta,\bar\zeta)$ aren't really independent, more care is needed. First, we reduce the area integral into a contour integral:
\begin{align}
 \int_\Omega d\zeta d\bar\zeta\,e^{2i\Re(A\xi\cosh\zeta)} = \int_\Omega d\zeta\,e^{iA\xi\cosh\zeta}\,d\bar\zeta\,e^{i\bar A\bar\xi\cosh\bar\zeta} = \oint_{\del\Omega} F d\bar\zeta\,e^{i\bar A\bar\xi\cosh\bar\zeta} \ , \label{eq:zeta_contour}
\end{align}
where $F(\zeta)$ is the primitive function of $e^{iA\xi\cosh\zeta}$. The integral along the boundary contour $\del\Omega$ is counter-clockwise, and consists of four straight segments, as depicted in figure \ref{fig:zeta}. 
We begin by considering segments 1,3. At any two ``opposite'' points $\zeta = \gamma\pm \pi i$ along these segments, the function $e^{i\bar A\bar\xi\cosh\bar\zeta}$ takes the same value $e^{-i\bar A\bar\xi\cosh\gamma}$. This brings the integral over these segments into the form:
\begin{align}
 \left(\int_1 + \int_3\right)F d\bar\zeta\,e^{i\bar A\bar\xi\cosh\bar\zeta} = \int_{-\infty}^\infty d\gamma\,e^{-i\bar A\bar\xi\cosh\gamma} \left( F(\gamma-\pi i) - F(\gamma + \pi i)\right) \ . \label{eq:integral_13_raw}
\end{align}
The difference in $F$ values can now be converted back into an integral $e^{iA\xi\cosh\zeta}d\zeta$ along the vertical segment $(\gamma-\pi i,\gamma + \pi i)$. Since $e^{iA\xi\cosh\zeta}$ takes the same value at $\zeta=\gamma \pm \pi i$, we can shift this vertical segment horizontally (by adding horizontal segments that will cancel), turning it into $(-\pi i,\pi i)$. The integral \eqref{eq:integral_13_raw} then factorizes, as:
\begin{align}
 \left(\int_1 + \int_3\right)F d\bar\zeta\,e^{i\bar A\bar\xi\cosh\bar\zeta} = -i\int_{-\infty}^\infty d\gamma\,e^{-i\bar A\bar\xi\cosh\gamma} \int_{-\pi}^\pi d\phi\,e^{iA\xi\cos\phi} = -4\pi i K_0(i\bar A\bar\xi) J_0(A\xi) \ , \label{eq:integral_13}
\end{align}
where $J_n$ is again the Bessel function of the first kind, and $K_n$ is the modified Bessel function of the second kind. We now turn to segments 2,4 of the contour $\del\Omega$, where $\Re\zeta = \pm\infty$. At these respective segments, $\cosh\zeta\approx \frac{1}{2}e^{\pm\zeta}$ is very large in absolute value, so the primitive function of $e^{iA\xi\cosh\zeta}$ can be approximated as:
\begin{align}
 F(\zeta) \approx c_\pm \pm \frac{e^{iA\xi\cosh\zeta}}{iA\xi\cosh\zeta} \ , \label{eq:F_asymp}
\end{align} 
where $c_\pm$ are integration constants that will be different between the two segments. At $\Im\zeta = 0$, the numerator in \eqref{eq:F_asymp} has absolute value smaller than 1, while the denominator is very large. Therefore, at these points $F$ is given simply by $c_\pm$, and we can find the difference between these two values as:
\begin{align}
 c_+ - c_- = F(+\infty) - F(-\infty) = \int_{-\infty}^\infty d\gamma\,e^{iA\xi\cosh\gamma} = 2 K_0(-iA\xi) \ .
\end{align}
At general points on segments 2,4, the integrand in \eqref{eq:zeta_contour} becomes:
\begin{align}
 F(\zeta) e^{i\bar A\bar\xi\cosh\bar\zeta} \approx c_\pm e^{i\bar A\bar\xi\cosh\bar\zeta} \pm \frac{e^{2i\Re(A\xi\cosh\zeta)}}{iA\xi\cosh\zeta} \ .
\end{align}
We can again neglect the second term, because the numerator has unit absolute value, while the denominator is very large. This brings the integrals over segments 2,4 into the form:
\begin{align}
 \left(\int_2 + \int_4\right)F d\bar\zeta\,e^{i\bar A\bar\xi\cosh\bar\zeta} = c_+ \int_2 d\bar\zeta\,e^{i\bar A\bar\xi\cosh\bar\zeta} + c_- \int_4 d\bar\zeta\,e^{i\bar A\bar\xi\cosh\bar\zeta} \ .
\end{align}
The vertical integration segments can now again be shifted horizontally, into the segment $(-\pi i,\pi i)$. This leads again to a factorized result:
\begin{align}
 \left(\int_2 + \int_4\right)F d\bar\zeta\,e^{i\bar A\bar\xi\cosh\bar\zeta} = -i(c_+ - c_-) \int_{-\pi}^\pi d\phi\,e^{i\bar A\bar\xi\cos\phi} = -4\pi i K_0(-iA\xi) J_0(\bar A\bar\xi) \ .
\end{align}
Altogether, the $d\zeta d\bar\zeta$ integral \eqref{eq:zeta_contour} evaluates into:
\begin{align}
 \int_\Omega d\zeta d\bar\zeta\,e^{2i\Re(A\xi\cosh\zeta)} = -4\pi i\left(J_0(A\xi)K_0(i\bar A\bar\xi) + J_0(\bar A\bar\xi)K_0(-iA\xi) \right) \ .
\end{align}
We can achieve a cleaner separation between $\xi$ and $\bar\xi$, using the fact that $\xi$ in \eqref{eq:I_big_integral} is always real or imaginary, and that $J_0$ is an even function:
\begin{align}
 \int_\Omega d\zeta d\bar\zeta\,e^{2i\Re(A\xi\cosh\zeta)} = -4\pi i\left(J_0(A\bar\xi)K_0(i\bar A\bar\xi) + J_0(\bar A\xi)K_0(-iA\xi) \right) \ . \label{eq:zeta_result}
\end{align}
This brings the inversion kernel \eqref{eq:I_big_integral} into the form:
\begin{align}
 \bbI_2(P_\mu,\tilde P_\mu) = \frac{1}{4\pi^3}\left( \int_0^\infty + \int_{i\infty}^0\right)\left[\xi d\xi J_0(\bar A\xi)K_0(-iA\xi) + \bar\xi d\bar\xi J_0(A\bar\xi)K_0(i\bar A\bar\xi) \right] \int_{-\infty}^\infty d\beta \ , \label{eq:I_big_integral_intermediate}
\end{align}
where we again used the fact that $\xi$ is either real or imaginary to replace $\xi d\xi$ by $\bar\xi d\bar\xi$ where convenient. We now see that, while the $\beta$ integral diverges, the $\xi,\bar\xi$ integral vanishes: the integrand has no singularities in the upper-right quadrant of the complex $\xi$ plane, and the contour can be closed at infinity. To extract an overall finite answer, we will employ a regularization that ties both integrals together. Notice that the divergence of the $\beta$ integral is a reflection of the infinite volume of the $SL(2,\bbR)$ residual symmetry group. We can thus regularize by imposing an ``IR cutoff'' within the $SL(2,\bbR)$. This can be done in a group-invariant fashion by imposing a cutoff $\Lambda$ on the \emph{trace} of the $SL(2,\bbR)$ matrix:
\begin{align}
 |\tr g| < \Lambda \ .
\end{align}
This will result in a cutoff $|\beta|<\beta_{\text{max}}$ on the $\beta$ integral, where $\beta_{\text{max}}$ depends on both $\Lambda$ and $\xi$; we assume an order of limits such that $\beta_{\text{max}}\gg 1$ everywhere. Explicitly, in the different integration domains \eqref{eq:g_spacelike_III},\eqref{eq:xi}, $\beta_{\text{max}}$ is given by:
\begin{align}
 \begin{split}
   \beta_{\text{max}} &= \left\{\begin{array}{ll}
      \ln\frac{\Lambda}{\cosh\theta} & \quad \text{Domain 1} \\  
      \ln\frac{\Lambda}{\cos\theta} & \quad \text{Domain 2} \\
      \ln\frac{\Lambda}{\sinh\theta} & \quad \text{Domain 3}
    \end{array} \right. \\
    &= \ln\Lambda - \frac{1}{2}\left\{\begin{array}{ll}
            \ln(\xi^2 + 1) & \quad \text{Domains 1,2} \\  
           \ln(-\xi^2 - 1) & \quad \text{Domain 3}
        \end{array} \right. \\
    &= \ln\Lambda - \frac{1}{2}\ln(\xi^2 + 1) + \left\{\begin{array}{ll}
      0 & \quad \text{Domains 1,2} \\  
      \frac{\pi i}{2} & \quad \text{Domain 3}
    \end{array} \right. \ .
 \end{split} \label{eq:beta_max}
\end{align}
Here, we used the standard definition of the complex logarithm, which has imaginary part $+\pi i$ when evaluated on negative numbers. With this definition, $\ln(\xi^2+1)$ has no singularities in the upper-right quadrant of the complex $\xi$ plane. In terms of $\bar\xi$, we can write instead:
\begin{align}
 \beta_{\text{max}} = \ln\Lambda - \frac{1}{2}\ln^*(\bar\xi^2 + 1) - \left\{\begin{array}{ll}
   0 & \quad \text{Domains 1,2} \\  
   \frac{\pi i}{2} & \quad \text{Domain 3}
 \end{array} \right. \ . \ , \label{eq:beta_max_bar}
\end{align}
where $\ln^*$ is an alternative branch of the logarithm, with imaginary part $-\pi i$ when evaluated on negative numbers. With this definition, $\ln(\bar\xi^2+1)$ again has no singularities in the upper-right quadrant of the complex $\xi$ plane. We now plug \eqref{eq:beta_max}-\eqref{eq:beta_max_bar} into \eqref{eq:I_big_integral_intermediate}, by changing the $\beta$ integration limits from $\pm\infty$ to $\pm\beta_{\text{max}}$. This yields:
\begin{align}
 &\bbI_2(P_\mu,\tilde P_\mu) = \frac{\ln\Lambda}{2\pi^3}\left(\int_0^\infty + \int_{i\infty}^0\right)\left[\xi d\xi J_0(\bar A\xi)K_0(-iA\xi) + \bar\xi d\bar\xi J_0(A\bar\xi)K_0(i\bar A\bar\xi) \right] \nonumber \\
   &\quad -\frac{1}{4\pi^3}\left(\int_0^\infty + \int_{i\infty}^0\right)\left[\xi d\xi J_0(\bar A\xi)K_0(-iA\xi)\ln(\xi^2 + 1) + \bar\xi d\bar\xi J_0(A\bar\xi)K_0(i\bar A\bar\xi)\ln^*(\bar\xi^2 + 1) \right] \nonumber \\
   &\quad +\frac{i}{4\pi^2}\int_{i\infty}^i \left[\xi d\xi J_0(\bar A\xi)K_0(-iA\xi) - \bar\xi d\bar\xi J_0(A\bar\xi)K_0(i\bar A\bar\xi) \right] \ . \label{eq:I_big_integral_regularized}
\end{align}
The first and second lines again vanish, because the integrand is regular in the upper-right quadrant, and the contour can be closed at infinity. We are left with the finite term on the third line, which evaluates to (setting $\xi\equiv ix$):
\begin{align}
 \begin{split}
   \bbI_2(P_\mu,\tilde P_\mu) &= \frac{1}{2\pi^2}\Im\int_1^\infty xdx\,I_0(Ax)K_0(\bar Ax) \\
    &= \frac{1}{2\pi^2}\Im\left.\frac{Ax I_1(Ax) K_0(\bar Ax) + \bar A x K_1(\bar Ax)I_0(Ax)}{A^2 - \bar A^2}\right|_1^\infty \\
    &= \frac{\Re\!\left[A\!\left(I_1(A) K_0(\bar A) + K_1(A)I_0(\bar A)\right)\right]}{4\pi^2\Im A^2} \ .
 \end{split}
\end{align}
Here, $I_n(x) = i^{-n} J_n(ix)$ is the modified Bessel function of the first kind, and $A$ is the function of $P_\mu,\tilde P_\mu$ given by \eqref{eq:A_spacelike_III}.

\subsection{Summary} \label{sec:inversion:summary}

In this section, we calculated two kernels for implementing inversions in Minkowski spacetime: $\bbI_1(p_\mu,\tilde p_\mu)$ for lightlike 4-momenta (or, equivalently, spatial 3-momenta), and $\bbI_2(P_\mu,\tilde P_\mu)$ for general 4-momenta. We defined these as the kernels for the decomposition of an inverted plane wave (with appropriate conformal weight) into ordinary plane waves:
\begin{align}
 \frac{e^{i\tilde p_\mu x^\mu/(x_\nu x^\nu)}}{x_\nu x^\nu} &= \int_{\substack{p^2\,=\,0 \\ p\cdot\tilde p\,<\,0}} \frac{d^3\mathbf{p}}{2|\mathbf{p}|}\,\bbI_1(p_\mu,\tilde p_\mu)\,e^{ip_\mu x^\mu} \ ; \label{eq:I_1_summary} \\
 \frac{e^{i\tilde P_\mu x^\mu/(x_\nu x^\nu)}}{(x_\nu x^\nu)^2} &= \int d^4P\,\bbI_2(P_\mu,\tilde P_\mu)\,e^{iP_\mu x^\mu} \ . \label{eq:I_2_summary}
\end{align}
For $\bbI_1(p_\mu,\tilde p_\mu)$, we found:
\begin{align}
 \bbI_1(p_\mu,\tilde p_\mu) = \frac{1}{2\pi}J_0\!\left(\!\sqrt{-2p_\mu \tilde p^\mu} \right) \ . \label{eq:I_1_summary_result}
\end{align}
The results for $\bbI_2(P_\mu,\tilde P_\mu)$ can be summarized as follows. $\bbI_2(P_\mu,\tilde P_\mu)$ is nonzero only when $P_\mu$ and $\tilde P_\mu$ have the same causal character (both timelike with the same time orientation, or both spacelike). In the timelike case, $\bbI_2(P_\mu,\tilde P_\mu)$ is given by:
\begin{align}
 \bbI_2(P_\mu,\tilde P_\mu) = \frac{A_+ J_1(A_+) J_0(A_-) - A_- J_1(A_-) J_0(A_+)}{4\pi(A_+^2 - A_-^2)} \ , \label{eq:I_2_summary_timelike}
\end{align}
while in the spacelike case, it's given by:
\begin{align}
 \bbI_2(P_\mu,\tilde P_\mu) = \begin{dcases}
    0 & \quad P_\mu\tilde P^\mu > m\tilde m \\
    -\frac{A_+ J_1(A_+) J_0(A_-) - A_- J_1(A_-) J_0(A_+)}{4\pi(A_+^2 - A_-^2)} & \quad P_\mu\tilde P^\mu < -m\tilde m \\  
    \frac{\Re\left[A\left(I_1(A) K_0(\bar A) + K_1(A)I_0(\bar A)\right)\right]}{4\pi^2\Im A^2} & \quad |P_\mu\tilde P^\mu| < m\tilde m
  \end{dcases} \ , \label{eq:I_2_summary_spacelike}
\end{align}
where we defined the shorthands:
\begin{align}
  m\tilde m &= \sqrt{(P_\mu P^\mu)(\tilde P_\nu\tilde P^\nu)} \ ; \\
  A_\pm &= \frac{1}{\sqrt{2}}\left( \sqrt{-P_\mu\tilde P^\mu + m\tilde m} \pm \sqrt{-P_\mu\tilde P^\mu - m\tilde m} \right) \ ; \label{eq:A_pm} \\
  A &= \frac{1}{\sqrt{2}}\left( \sqrt{m\tilde m + P_\mu\tilde P^\mu} + i\,\sqrt{m\tilde m - P_\mu\tilde P^\mu} \right) \ . \label{eq:A}
\end{align}
We note that both inversion kernels are symmetric in their arguments, i.e. $\bbI_1(p_\mu,\tilde p_\mu) = \bbI_1(\tilde p_\mu,p_\mu)$ and $\bbI_2(P_\mu,\tilde P_\mu) = \bbI_2(\tilde P_\mu,P_\mu)$.

\subsection{Consistency check and the near-lightlike limit} \label{sec:inversion:check}

Having obtained results for the inversion of lightlike and non-lightlike momenta, it is interesting to compare the two. Consider first a plane wave with \emph{timelike} 4-momentum $\tilde P_\mu$, which is close to being lightlike (in some preferred reference frame). What are the coefficients $\bbI_2(P_\mu,\tilde P_\mu)$ for its decomposition into new timelike momenta $P_\mu$ upon inversion? Plugging $\tilde m=0$ into \eqref{eq:I_2_summary_timelike},\eqref{eq:A_pm}, we get:
\begin{align}
 &A_+ = \sqrt{-2P_\mu\tilde P^\mu} \ ; \quad A_- = 0 \label{eq:A_pm_null} \\
 &\quad \Longrightarrow \quad \bbI_2(P_\mu,\tilde P_\mu) = \frac{1}{4\pi\sqrt{-2P_\mu\tilde P^\mu}}\,J_1\!\left(\sqrt{-2P_\mu\tilde P^\mu}\right) \ . \label{eq:I_timelike_null}
\end{align}
In particular, even though $\tilde P_\mu$ is almost lightlike, \eqref{eq:I_timelike_null} is not exclusively concentrated at almost-lightlike $P_\mu$. This is not as strange as it may seem: while the lightcone is invariant under inversion, the property of being ``almost lightlike'' is not! We can similarly consider a \emph{spacelike} 4-momentum $\tilde P_\mu$ that's close to being lightlike. In Region II of \eqref{eq:I_2_summary_spacelike}, we again obtain eqs. \eqref{eq:A_pm_null}-\eqref{eq:I_timelike_null}, but with an opposite overall sign in \eqref{eq:I_timelike_null}:
\begin{align}
 \bbI_2(P_\mu,\tilde P_\mu) = -\frac{1}{4\pi\sqrt{-2P_\mu\tilde P^\mu}}\,J_1\!\left(\sqrt{-2P_\mu\tilde P^\mu}\right) \ . \label{eq:I_spacelike_null_II}
\end{align}
As for Region III of \eqref{eq:I_2_summary_spacelike}, it shrinks in the limit of near-lightlike $\tilde P_\mu$, and the function $A$ from \eqref{eq:A_spacelike_III},\eqref{eq:A} becomes small there: 
\begin{align}
 A = \sqrt{m\tilde m}\,e^{i\chi/2} \, \rightarrow \, 0 \ .
\end{align}
This makes $\bbI_2(P_\mu,\tilde P_\mu)$ in Region III \emph{large}. In particular, the second term in the numerator in \eqref{eq:I_2_summary_spacelike} dominates, and we get:
\begin{align}
 \bbI_2(P_\mu,\tilde P_\mu) = \frac{1}{4\pi^2\Im A^2} = \frac{1}{4\pi^2\sqrt{(P_\mu P^\mu)(\tilde P_\mu\tilde P^\mu) - (P_\mu\tilde P^\mu)^2}} \ . \label{eq:I_spacelike_null_III}
\end{align}
At any rate, we see that simply taking a lightlike limit does not allow for a direct comparison between the lightlike and non-lightlike inversion formulas. However, we can make a less direct comparison, which will serve as a strong consistency check on this section's results. Let's evaluate the non-lightlike inversion formula \eqref{eq:I_2_summary} at $t=0$:
\begin{align}
 \frac{e^{i\mathbf{\tilde P\cdot x}/\mathbf{x}^2}}{\mathbf{x}^4} = \int d^4P\,\bbI_2(P_\mu,\tilde P_\mu)\,e^{i\mathbf{P\cdot x}} \ , \label{eq:I_2_spatial}
\end{align}
and compare this with the $t$ derivative of the lightlike formula \eqref{eq:I_1_summary}, also at $t=0$:
\begin{align}
 \frac{i|\mathbf{\tilde p}|e^{i\mathbf{\tilde p\cdot x}/\mathbf{x}^2}}{\mathbf{x}^4} = \frac{i}{2}\int d^3\mathbf{p}\,\bbI_1(\mathbf{p},\mathbf{\tilde p})\,e^{i\mathbf{p\cdot x}} \ . \label{eq:I_1_spatial}
\end{align}
Here, we assumed positive $\tilde p_t$, i.e. $\tilde p_t = +|\mathbf{\tilde p}|$, without loss of generality. Comparing eqs. \eqref{eq:I_2_spatial}-\eqref{eq:I_1_spatial}, we conclude that the two inversion kernels must be related by:
\begin{align}
 \int dP_t\,\bbI_2(P_\mu,\tilde P_\mu) = \frac{\bbI_1(\mathbf{P},\mathbf{\tilde P})}{2|\mathbf{\tilde P}|} \ , \label{eq:consistency}
\end{align}
for \emph{any} choice of time component $\tilde P_t$ on the LHS. Without loss of generality, let us choose $\tilde P_t>0$. Then, for $\tilde P_t>|\mathbf{\tilde P}|$, the integral in \eqref{eq:consistency} probes the timelike regime \eqref{eq:I_2_summary_timelike}, and ranges over $P_t\in(|\mathbf{P}|,\infty)$. For $\tilde P_t< |\mathbf{\tilde P}|$, the integral probes instead the spacelike regime \eqref{eq:I_2_summary_spacelike}, where we encounter two possible situations. If $\tilde P_t < \mathbf{\tilde P\cdot P}/|\mathbf{P}|$, then the integral captures only Region III of \eqref{eq:I_2_summary_spacelike}, with integration limits $P_t\in(E_-,E_+)$ given by:
\begin{align}
 E_\pm = \frac{1}{\mathbf{\tilde P}^2}\left(\mathbf{P\cdot\tilde P}\,\tilde P_t \pm \sqrt{\left(\mathbf{\tilde P}^2 - (\tilde P_t)^2 \right) \left(\mathbf{P}^2\mathbf{\tilde P}^2 - (\mathbf{P\cdot\tilde  P})^2 \right)} \right) \ . \label{eq:E_pm}
\end{align}
If, on the other hand, $\tilde P_t > \mathbf{\tilde P\cdot P}/|\mathbf{P}|$, then \emph{in addition} to this range in Region III, we also have the range $P_t\in(E_+,|\mathbf{P}|)$ that lies in Region II of \eqref{eq:I_2_summary_spacelike}.

For all these different cases, we verified that the consistency relation \eqref{eq:consistency} holds, via numerical integration with various choices of the parameters. In the near-lightlike limit discussed above, we were also able to perform this check analytically. In the spacelike case, this requires taking into account both the Region II contribution, where the integrand is given by \eqref{eq:I_spacelike_null_II}, and the Region III contribution, where the integrand is given by \eqref{eq:I_spacelike_null_III}. In the latter, the very small integration range and the very large integrand combine into a finite contribution.

\section{Putting together the static-patch scattering result} \label{sec:result}

We now return to de Sitter space, and to our scattering formulas \eqref{eq:S_general},\eqref{eq:S2_integrated}, where both the Poincare-patch evolution function $\hat M_2$ and the inversion kernels $\bbI_{1,2}$ are now known. For convenience, let us reproduce here the relevant formulas:
\begin{align}
 \begin{split}
  a_H(k_\mu) ={}& \int_{\substack{\tilde k^2\,=\,0 \\ k\cdot\tilde k\,<\,0}} \frac{d^3\mathbf{\tilde k}}{2|\mathbf{\tilde k}|}\,\bbI_1(k_\mu,\tilde k_\mu)\,\tilde a_{\tilde H}(\tilde k_\mu) \\
   &+ \int_{\tilde p^2 = 0} \frac{d^3\mathbf{\tilde p}}{2|\mathbf{\tilde p}|} \int_{\tilde q^2 = 0} \frac{d^3\mathbf{\tilde q}}{2|\mathbf{\tilde q}|}\,S_2(k_\mu;\tilde p_\mu+\tilde q_\mu)\,\tilde a_{\tilde H}(\tilde p_\mu)\,\tilde a_{\tilde H}(\tilde q_\mu) \ ;
 \end{split} \label{eq:S_general_summary} \\
 \begin{split}
  S_2(k_\mu;\tilde P_\mu) ={}& -\left.\bbI_1(k_\mu,\tilde k_\mu)\,\hat M_2(\tilde k_\mu;\tilde P_\mu)\right|_{\tilde k_\mu\,=\,(\sign(k_t)|\mathbf{\tilde P}|,\,\mathbf{\tilde P})} \\
   &+ \int dP_t\left.\bbI_2(P_\mu,\tilde P_\mu)\,\hat M_2(k_\mu;P_\mu)\right|_{\mathbf{P} = \mathbf{k}} \ ;
 \end{split} \label{eq:S_summary} \\
 \hat M_2(k_\mu;P_\mu) ={}& i\alpha\sign(k_t)\left(\frac{1}{\varepsilon} - \ln|P_t - k_t| + \frac{\pi i}{2}\sign(P_t - k_t) \right) \ . \label{eq:M_summary}
\end{align}
Now, let's consider the fate of the three terms in the Poincare-patch evolution \eqref{eq:M_summary}, paying attention to antipodal symmetry $t\rightarrow -t$, i.e. to parity under the flipping of energy signs. The first term is divergent. As already discussed, we remove it by restricting to even incoming data, $a_{\tilde H}(-|\mathbf{\tilde p}|,\mathbf{\tilde p}) = -a_{\tilde H}(|\mathbf{\tilde p}|,\mathbf{\tilde p}) \equiv \frac{i}{2}\,c_{\tilde H}(\mathbf{\tilde p})$. We then decompose the final data $a_H(k_\mu)$ into odd and even parts $b_H(\mathbf{k}),c_H(\mathbf{k})$. The second term in \eqref{eq:M_summary} will contribute only to the even part, while the third term will contribute only to the odd part. Furthermore, this latter contribution to the odd part actually \emph{vanishes}, due to a cancellation between the two terms in \eqref{eq:S_summary}. Indeed, inside the $P_t$ integral in \eqref{eq:S_summary}, $\sign(P_t - k_t)$ is always a constant, and the cancellation then follows from the identity \eqref{eq:consistency}. This leaves us only with even final data $c_H(\mathbf{k})$, coming from the second term in \eqref{eq:M_summary}. Thus, our final result consists of scattering from even data on the initial horizon to even data on the final horizon:
\begin{align}
 \begin{split}
   c_H(\mathbf{k}) ={}& \int \frac{d^3\mathbf{\tilde k}}{2|\mathbf{\tilde k}|}\,\bbI_1(\mathbf{k},\mathbf{\tilde k})\,\tilde c_{\tilde H}(\mathbf{\tilde k})
    + \int \frac{d^3\mathbf{\tilde p}}{2|\mathbf{\tilde p}|} \int \frac{d^3\mathbf{\tilde q}}{2|\mathbf{\tilde q}|}\,S_{\text{even}}(\mathbf{k};\mathbf{\tilde p}, \mathbf{\tilde q})\, \tilde c_{\tilde H}(\mathbf{\tilde p})\,\tilde c_{\tilde H}(\mathbf{\tilde q}) \ ,
 \end{split} \label{eq:S_even_general}
\end{align}
where the scattering kernel is given by:
\begin{align}
 \begin{split}
   S_{\text{even}}(\mathbf{k};\mathbf{\tilde p}, \mathbf{\tilde q}) = \frac{\alpha}{2}&\left(\bbI_1(\mathbf{k},\mathbf{\tilde p+\tilde q})\,
    \ln\frac{(|\mathbf{\tilde p}| + |\mathbf{\tilde q}| + |\mathbf{\tilde p+\tilde q}|)(|\mathbf{\tilde p}| + |\mathbf{\tilde q}| - |\mathbf{\tilde p+\tilde q}|)}
               {(-|\mathbf{\tilde p}| + |\mathbf{\tilde q}| + |\mathbf{\tilde p+\tilde q}|)(|\mathbf{\tilde p}| - |\mathbf{\tilde q}| + |\mathbf{\tilde p+\tilde q}|)} \right. \\
    &\quad - \int_{|\mathbf{k}|}^\infty dP_t\,\ln(P_t^2 - \mathbf{k}^2)\left.\bbI_2(P_\mu,\tilde P_\mu)\right|_{\mathbf{P} = \mathbf{k};\,\mathbf{\tilde P} = \mathbf{\tilde p + \tilde q};\,\tilde P_t = |\mathbf{\tilde p}| + |\mathbf{\tilde q}|} \\
    &\quad + \left.\int_{-|\mathbf{k}|}^{|\mathbf{k}|} dP_t\,\ln(\mathbf{k}^2 - P_t^2)\left.\bbI_2(P_\mu,\tilde P_\mu)\right|_{\mathbf{P} = \mathbf{k};\,\mathbf{\tilde P} = \mathbf{\tilde p + \tilde q};\,\tilde P_t = |\mathbf{\tilde p}| - |\mathbf{\tilde q}|} \ \right) \ .
 \end{split} \label{eq:S_even}
\end{align}
Here, the second line carries the contribution to \eqref{eq:S_general_summary} from $\tilde p_\mu,\tilde q_\mu$ of the same energy sign, such that $\tilde P_\mu = \tilde p_\mu+\tilde q_\mu$ is timelike, while the third line carries the contribution from $\tilde p_\mu,\tilde q_\mu$ of opposite energy signs, such that $\tilde P_\mu = \tilde p_\mu+\tilde q_\mu$ is spacelike. The limits of the $P_t$ integrals and the corresponding regimes of $\bbI_2(P_\mu,\tilde P_\mu)$ are the same as in the paragraph following \eqref{eq:consistency}, except that we now allow $\tilde P_t<0$ in the spacelike case. In particular, the second line of \eqref{eq:S_even} is probing the timelike regime \eqref{eq:I_2_summary_timelike}, while the third line is probing the spacelike regime \eqref{eq:I_2_summary_spacelike}. The integration limits in the third line are governed by $E_\pm$ from \eqref{eq:E_pm}; in the range $P_t\in (E_-,E_+)$, the integral probes Region III of \eqref{eq:I_2_summary_spacelike}; if $\mathbf{k}^2 (|\mathbf{\tilde p}| - |\mathbf{\tilde q}|)^2 > (\mathbf{k\cdot (\tilde p + \tilde q)})^2$, then Region II of \eqref{eq:I_2_summary_spacelike} is probed as well, in the range $P_t\in (E_+,|\mathbf{k}|)$ if $\tilde P_t = |\mathbf{\tilde p}| - |\mathbf{\tilde q}|$ is positive, or $P_t\in (-|\mathbf{k}|,E_-)$ if it is negative. The remainder of the $(-|\mathbf{k}|,|\mathbf{k}|)$ integration range in \eqref{eq:S_even} probes Region I of \eqref{eq:I_2_summary_spacelike}, where $\bbI_2$ vanishes.

Let us briefly comment on the behavior of our answer \eqref{eq:S_even} under 3d dilatations, i.e. under time translations in the static patch. The Poincare-patch evolution \eqref{eq:M_summary} contains a logarithm, which transforms inhomogeneously under dilatations, via an additive constant. This is a symptom of the divergence contained in the $1/\varepsilon$ term. Having cancelled the divergence by restricting to even initial data, we should find that the inhomogeneous transformations of the logarithms in \eqref{eq:S_even} cancel as well. In the first line of \eqref{eq:S_even}, the cancellation is manifest, since the logarithm there is of a dimensionless ratio. This is not the case in the second and third lines. However, it's easy to see that the inhomogeneous transformations of the second and third lines cancel each other, as a result of the consistency relation \eqref{eq:consistency}.

\section{Discussion} \label{sec:discuss}

In this paper, we calculated the evolution of a conformally massless scalar field with cubic interaction from the initial horizon to the final horizon of a $dS_4$ static patch. To our knowledge, this is the first such calculation of the most natural observable in de Sitter space. The calculation proceeded by (1) extending the field data on the half-horizon boundaries of the static patch to antipodally even configurations on entire horizons, (2) evolving the data between each of the horizons and the unobservable conformal boundary of $dS_4$, and (3) sewing these two evolutions together by a coordinate inversion. The main technical difficulty was to work out the relevant inversion formulas in momentum variables. These formulas can be defined in terms of flat spacetime, and should be useful in a broader context. 

Our final result \eqref{eq:S_even} for the static-patch scattering is not quite given in closed form, since we haven't managed to perform analytically the energy integral $\int dP_t$. It would be interesting to do so, if not in general then at least in some limits. In any case, we stress that our ability to reduce things to this single integral relied on the simplicity of working with a single cubic vertex. For diagrams with quartic vertices, or with more than one vertex, the general procedure from eqs. \eqref{eq:reduction_1}-\eqref{eq:reduction_2} will still be valid, but it's no longer clear how to reduce the inversions to just a single integral. Perhaps a better way forward would be to mimic the modern flat-spacetime scattering industry (see e.g. \cite{Dixon:1996wi,Britto:2004ap,Britto:2004nc}) and its relatives in inflationary cosmology (e.g. \cite{Arkani-Hamed:2018kmz}), and try to reconstruct higher-point functions from the cubic one using some general principles, instead of resorting again to a bulk calculation. 

A more straightforward next step is to repeat the cubic-vertex calculation for massless theories with spin, i.e. Yang-Mills and General Relativity. The techniques developed here should be applicable with some slight modifications, especially since we already relied on spinors for the inversion calculation.  

In closing, let's return to the subject of cancellations between the two Poincare-patch evolutions in eq. \eqref{eq:reduction_2}. As discussed there, such a cancellation will occur whenever the Poincare-patch evolution (or a piece thereof) has the full $SO(1,4)$ de Sitter symmetry, i.e. the full conformal symmetry on the boundary. For processes with a single interaction vertex, this is always \emph{almost} the case. This is because the $dS_4$ Poincare-patch calculation is very closely related to the Euclidean $AdS_4$ boundary correlator in Poincare coordinates, which is $SO(1,4)$-invariant. The difference that can (and does) spoil the invariance is that our static-patch calculation requires both positive and negative Poincare-patch energies, whereas in the Euclidean calculation the energies all have the same sign. Flipping the energy signs can then introduce singularities in momentum space that weren't present in the Euclidean picture, such as when $1/(|\mathbf{P}| + |\mathbf{k}|)$ turns into $1/(|\mathbf{P}| - |\mathbf{k}|)$. Applying a special conformal generator, we then find a non-vanishing contribution from these singularities. However, the converse is also true: whenever a Euclidean correlator (or a piece thereof) does \emph{not} have singularities when continued to opposite energy signs, we should expect $SO(1,4)$-invariance of the Poincare-patch evolution, and a cancellation in the static-patch scattering. In the present paper, we've seen this in the case of the third term in eq. \eqref{eq:M_summary}. Though we did not describe that cancellation as a consequence of symmetry in the main text, it is in fact reflecting the conformal symmetry of the pure contact term $\delta^3(\mathbf{P-k})$. For Yang-Mills and GR, we expect similar cancellations for static-patch scattering with $(+++)$ and $(---)$ helicities, since the Euclidean boundary correlators for these helicity choices don't have the energy poles that are present for helicities $(++-)$ and $(--+)$ \cite{Maldacena:2011nz}. Such a cancellation would make for a nice similarity between $dS_4$ static-patch scattering and its more symmetric counterpart, the Minkowski S-matrix.

Another complication is that the $SO(1,4)$ symmetry may be already broken in the Euclidean correlators, due to a bulk IR divergence. This happened in our present case, producing the logarithmic term in the Poincare-patch evolution \eqref{eq:M_summary}. In Yang-Mills theory, this issue should not arise. For GR, the question is more subtle, as e.g. in \cite{Maldacena:2002vr} there's a divergence that ends up cancelling between the backwards and forwards Poincare-patch evolutions. In our setup, there will be an inversion between these two evolutions, which may spoil the cancellation. We will then be forced to choose, as in the present paper, initial data in a combination that falls off sufficiently quickly at $\scri$. 

One of our long-term goals in this work is to build towards bulk de Sitter observables (such as static-patch scattering amplitudes) for higher-spin gravity \cite{Vasiliev:1995dn,Vasiliev:1999ba}, and in particular for higher-spin dS/CFT \cite{Anninos:2011ui}. It is our hope that the study of more ordinary interacting massless theories can provide some guidance in this direction.

\section*{Acknowledgements}

We are grateful to Adrian David, Sudip Ghosh, Mirian Tsulaia, Slava Lysov, David O'Connell and Miquel Jorquera for discussions. YN's thinking was substantially informed by his participation in the CERN virtual workshop ``Cosmological correlators''. In particular, comments by Guilherme Pimentel and Nima Arkani-Hamed on an early version of this project have been instrumental. This work was supported by the Quantum Gravity Unit of the Okinawa Institute of Science and Technology Graduate University (OIST), which hosted EA as an intern during the project's early stages.

\end{document}